%% file: main.tex
\documentclass[11pt]{article}
\usepackage[utf8]{inputenc}
\usepackage{amsmath, amssymb, enumerate, graphicx, fullpage, physics, bm, upgreek}

\usepackage{appendix}
\usepackage{natbib}
\usepackage{sidecap}
\usepackage{titling}
\usepackage{setspace}
\usepackage[margin=1in]{geometry}
\usepackage{overpic, rotating, pbox}

\usepackage{palatino} 

\newcommand{\E}{\mathop{\mathbb{E}}}
\newcommand{\R}{\mathbb{R}}

\newcommand\Rey{\mbox{\textit{Re}}}  
\newcommand\Real{\mbox{Re}}          
\newcommand\Imag{\mbox{Im}}          

\DeclareOldFontCommand{\rm}{\normalfont\rmfamily}{\mathrm}
\DeclareOldFontCommand{\sf}{\normalfont\sffamily}{\mathsf}
\DeclareOldFontCommand{\tt}{\normalfont\ttfamily}{\mathtt}
\DeclareOldFontCommand{\bf}{\normalfont\bfseries}{\mathbf}
\DeclareOldFontCommand{\it}{\normalfont\itshape}{\mathit}
\DeclareOldFontCommand{\sl}{\normalfont\slshape}{\@nomath\sl}
\DeclareOldFontCommand{\sc}{\normalfont\scshape}{\@nomath\sc}
\DeclareRobustCommand*\cal{\@fontswitch\relax\mathcal}
\DeclareRobustCommand*\mit{\@fontswitch\relax\mathnormal}

\IfFileExists{t1phv.fd}
{\DeclareTextFontCommand\textsfbi{\usefont{T1}{phv}{b}{it}}
\DeclareMathAlphabet\mathsfbi            {T1}{phv}{b}{it}
}{}

\DeclareMathAlphabet{\mathcal}{OMS}{cmsy}{m}{n}
\DeclareMathAlphabet\mathbfcal{OMS}{cmsy}{b}{n}

\usepackage{xcolor}

\title{\fontsize{19}{19}{\textbf{Multiscale model reduction for incompressible flows}}}

\author{\normalsize{Jared L. Callaham$^{1*}$, Jean-Christophe Loiseau$^2$, and Steven L. Brunton$^1$}\\
	\footnotesize{$^1$ Department of Mechanical Engineering, University of Washington,
    Seattle, WA 98195, USA}\\
    \footnotesize{
	$^2$ Arts et M\'{e}tiers Institute of Technology, CNAM, DynFluid, HESAM Universit\'{e}, F-75013 Paris, France
	}
	}

\begin{document}
\date{}
\maketitle
\vspace{-.2in}
\begin{abstract}
Many unsteady flows exhibiting complex dynamics are nevertheless characterized by emergent large-scale coherence in space and time.
Reduced-order models based on Galerkin projection of the governing equations onto an orthogonal modal basis approximate the flow as a low-dimensional dynamical system with linear and quadratic terms.
However, these Galerkin models often fail to reproduce the true dynamics, in part because they ignore important nonlinear interactions with unresolved flow scales. 
Here, we use a separation of time scales between the resolved and subscale variables to derive a reduced-order model with cubic closure terms for the truncated modes, generalizing the classic Stuart-Landau equation.  
The leading-order cubic terms are determined by averaging out fast variables through a perturbation series approximation of the action of a stochastic Koopman operator.  
We show analytically that this multiscale closure model can capture both the effects of mean-flow deformation and the energy cascade before demonstrating improved stability and accuracy in models of chaotic lid-driven cavity flow and vortex pairing in a mixing layer.
This approach to closure modeling establishes a general theory for the origin and role of cubic nonlinearities in low-dimensional models of incompressible flows.\\

\end{abstract}

\input{S1_Introduction.tex}

\input{S2_Galerkin.tex}

\input{S3_Generators.tex}

\input{S4_Closure.tex}

\input{S5_Numerics.tex}

\input{S6_Results.tex}

\input{S7_Discussion.tex}

\section*{Acknowledgements}
SLB acknowledges funding support from the Army Research Office (ARO W911NF-19-1-0045) and National Science Foundation AI Institute in Dynamic Systems (grant number 2112085).
JLC acknowledges funding support from the Department of Defense (DoD) through the National Defense Science \& Engineering Graduate (NDSEG) Fellowship Program. 

\section*{Declaration of interests}
The authors report no conflict of interest.

\appendix
\input{A1_Stability.tex}

\input{A2_Truncation.tex}

\input{A3_Closures.tex}

\clearpage
\setlength{\bibsep}{2.4pt plus 1ex}
\begin{spacing}{.01}
	\small
	\bibliographystyle{jfm}
	\bibliography{refs}
\end{spacing}

\end{document}

%% file: S1_Introduction.tex
\section{Introduction}
\label{sec: introduction}

Despite the complex and often chaotic dynamics exhibited by many unsteady fluid flows, they are often dominated by energetic coherent structures evolving on relatively long length and time scales~\citep{Holmes1996}. 
This realization made it possible to study fluid flows as high-dimensional dynamical systems evolving on a low-dimensional manifold, providing answers to longstanding questions on topics such as the route to turbulence~\citep{Landau1944, Hopf1948, Ruelle1971cmp, SwinneyGollub} and the role of nonlinear interactions~\citep{Stuart1958jfm, Landau1959book}. 
The persistent nature of these coherent structures eventually also raised the possibility of using low-dimensional surrogate models for optimization and control objectives~\citep{Noack2011book, Brunton2015amr, Rowley2017}.

The fundamental challenges in constructing such reduced-order models may be broken into two categories: approximating the structures themselves, and approximating their evolution.
We refer to these as the kinematic and dynamic approximations, respectively.
A modal separation of variables assumption is often employed for the former task~\citep{Taira2017,Taira2020aiaa}, but although this may be suitable for many closed or diffusion-dominated flows, it is not a natural representation of traveling waves or advection-dominated flows~\citep{Rowley2000physd, Reiss2018, Rim2018juq, Grimberg2020jcp, Mendible2020tcfd}.
This issue is inextricably linked to the problem of modeling the coherent structure dynamics; as is well known in many domains, a proper choice of coordinates can greatly simplify the modeling task~\citep{Champion2019pnas}.

The multiscale nature of fluid flows further complicates both the kinematic and dynamic aspects of low-dimensional modeling.
For instance, the effective dimensionality of a chaotic or turbulent flow may be orders of magnitude greater than that of a laminar flow with one or two instability modes.
Meanwhile, the ``triadic'' structure of the nonlinear interactions in the wavenumber or frequency domain ensures that the dynamics of all scales across the flow are linked.
Thus, even if the large, energetic coherent structures can be approximated with a low-dimensional basis, models of their dynamics that do not account for the role played by the unresolved degrees of freedom are often unstable or physically inconsistent~\citep{Noack2011book, Callaham2021langevin}.
Similar considerations impact the development of numerical methods~\citep{Bazilevs2007}, self-consistent mean flow modeling~\citep{Meliga2017jfm}, and resolvent analysis~\citep{Padovan2020jfm, Rigas2020jfm, Barthel2021jfm, Barthel2022prf}.

The need for subscale modeling was apparent even in early work combining empirical modal approximations, such as the proper orthogonal decomposition (POD), with physics-based model reduction, such as Galerkin projection.
For example, low-dimensional models of vortex shedding in the globally unstable cylinder wake could accurately predict the dynamics over short times, but were subject to structural instability over a longer time horizon~\citep{Deane1991pof, Ma2002jfm}.
Eventually~\citet{Noack2003jfm} showed that this was a result of the failure of the standard post-transient POD basis to resolve the Stuart-Landau mechanism of mean flow deformation associated with nonlinear interactions between the fluctuations~\citep{Landau1944, Stuart1958jfm}, an insight that enabled low-dimensional modeling of natural and actuated flows with increasingly more complex dynamics~\citep{Luchtenburg2009jfm, Deng2020jfm, Sieber2020, deng2021galerkin, Callaham2021wake}.

Contemporary work on POD-Galerkin modeling of turbulent shear flows also recognized the need for closure models that could approximate the effect of unresolved scales~\citep{Aubry1988jfm, Rempfer1994, Ukeiley2001}.
These efforts targeted two distinct physical mechanisms: mean flow deformation and subscale dissipation.
For flows that are either parallel~\citep{Aubry1988jfm} or weakly nonparallel under the assumption of Taylor's frozen turbulence hypothesis~\citep{Ukeiley2001}, these authors developed a Boussinesq eddy viscosity relationship between the resolved scales and a slowly varying parallel mean flow, leading to stabilizing cubic terms consistent with the Stuart-Landau description.
To capture dissipation due to the unresolved scales, these models also adopted a linear mixing length approximation.

Although these studies remain landmark explorations of low-dimensional coherent structure modeling, there are several opportunities to improve the proposed closure strategies.
First, it is difficult to generalize the Reynolds stress models applied to parallel shear flows by~\citet{Aubry1988jfm} and ~\citet{Ukeiley2001} to fully inhomogeneous flows.
While the ``shift mode'' approach to mean flow modeling via an augmented POD basis introduced by~\citet{Noack2003jfm} is agnostic to the geometry of the flow, it requires computation of the unstable steady state of the Navier-Stokes equations, which is not generally experimentally accessible.
Second, in the Richardson-Kolmogorov energy cascade description, energy is transferred from large to small scales through \emph{nonlinear} interactions, where it is finally dissipated.
This is not consistent with a linear mixing length model, a fact exploited by later work investigating nonlinear models of subscale dissipation~\citep{Wang2012cmame, Cordier2013, Osth2014jfm}, particularly the finite time thermodynamics approach, which is centered on modeling unresolved nonlinear energy transfers~\citep{Noack2008jnet}.

Recent years have seen a surge in interest in data-driven and machine learning-based methods~\citep{Brenner2019prf, Duraisamy2018arfm, Brunton2020arfm}, including a number of proposed closure and stabilization schemes for reduced-order models, either through regression to additional linear-quadratic terms~\citep{Xie2018siam, Mohebujjaman2017jcp, Mohebujjaman2018ijnmf} or by adding a deep learning model to approximate the residual~\citep{San2018acm, San2018extreme, Menier2022}.
Alternative work has explored interpretable system identification methods that forego the projection-based model altogether~\citep{Brunton2016pnas, Loiseau2017jfm, Peherstorfer2016cmame, Qian2020, Callaham2021wake}, but may incorporate physical constraints derived from the Galerkin system~\citep{Loiseau2018jfm, Deng2020jfm, Kaptanoglu2021}. 
If an accurate, nonintrusive model is more important than interpretability or satisfying physical constraints, then
the traditional projection-based framework can be eliminated altogether with black-box neural network forecasting methods~\citep{Hesthaven2018jcp, Wan2018plos}.

These empirical closure models are constructed on the assumption that the influence of the unresolved variables can be approximated based on information from the resolved variables alone.
The success of the Reynolds stress-mean flow models~\citep{Aubry1988jfm, Ukeiley2001, ManticLugo2014prl}, invariant or center manifold reductions~\citep{GuckenheimerHolmes, Coullet1983, Noack2003jfm, Carini2015jfm}, and weakly nonlinear analysis~\citep{Stuart1958jfm, Sipp2007jfm, Meliga2009jfm, Meliga2011jfm} suggests that there may be circumstances where this relationship may be derived analytically via traditional analysis.
The unifying thread between these methods is the assumption of a scale separation between the resolved and unresolved variables that can be exploited to develop an asymptotically correct closure model.

\begin{figure}
\centering
\includegraphics[width=\textwidth]{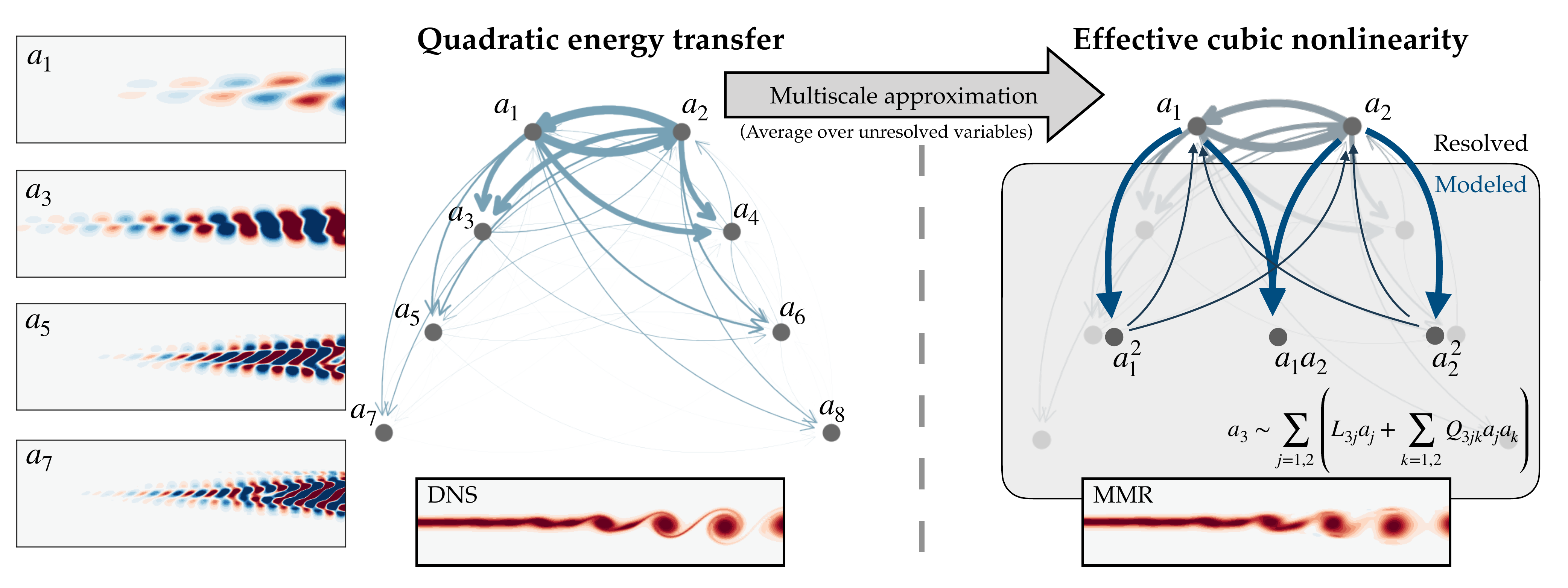}
\caption{
{\small
\textbf{Multiscale closure model applied to a mixing layer.}
The visualization of the network of average quadratic energy transfer between the leading harmonic modes shows the cascade of energy to higher-order modes.
The multiscale model reduction (MMR) method approximates the effects of unresolved higher-order modes via stochastic averaging, which leads to a generalized Stuart-Landau-type equation with cubic nonlinear interactions.
The network visualization of quadratic energy transfers (left) is computed from the modal coefficients and Galerkin model for the leading harmonics, while that of of the multiscale approximation (right) is a notional illustration of the origin of the cubic terms in Eq.~\eqref{eq: closure -- stuart-landau}.
See Sec.~\ref{sec: results -- mixing-layer} for details on the construction of this figure and the low-dimensional model of the mixing layer.
}
}
\label{fig: intro -- overview}
\end{figure}

Beyond the field of fluid dynamics, the method of adiabatic elimination~\citep{Haken1983, Risken1996book} has long been used to discard the fast variables in systems with emergent large-scale coherence when there is a separation in time scales, while heterogeneous multiscale methods have played an important role in simulating physical systems with widely separated scales~\citep{EWeinan2003, EWeinan2007, EWeinan2011book}.
A similar stochastic averaging approach has been successful in climate modeling~\citep{Majda2001cpam}, where the primitive equations have the same quadratic nonlinearity as the usual Navier-Stokes equations without rotation, buoyancy, topography, etc.
This method, also called homogenization in the multiscale modeling literature, has its roots in the theory of singular perturbations of Markov processes~\citep{Kurtz1973, Papanicolaou1976} and is rigorously supported for stochastic systems with asymptotic scale separation; see~\citet{Majda2001cpam, Givon2004nonlinearity, EWeinan2011book}, or~\citet{PavliotisStuart2012} for in-depth presentations.
With some assumptions on ergodicity, a similar approach can also be taken with deterministic systems, even in a regime where the scale separation is not in the asymptotic limit~\citep{Majda2006nonlinearity}.

\begin{figure}
\centering
\includegraphics[width=\textwidth]{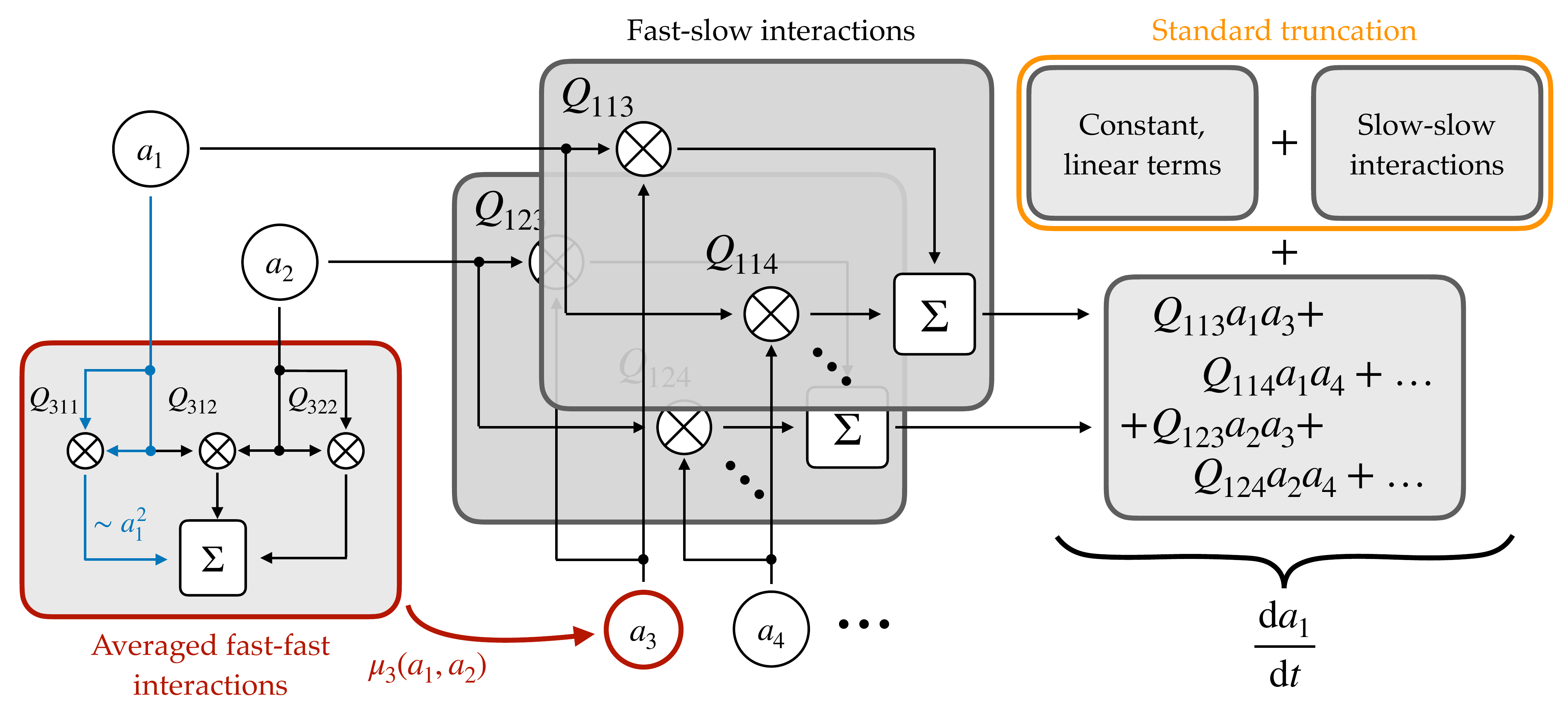}
\caption{
{\small
\textbf{Schematic of nonlinear interactions in the multiscale closure scheme.} 
The dynamics of one variable ($a_1$) in a system with two slow variables involve quadratic interactions between fast and slow variables that would be neglected in a standard truncation.
Instead, the proposed method averages over the fast scales, ultimately generating effective cubic nonlinearities in the closed equations.
}}
\label{fig: intro -- nonlinearity}
\end{figure}

This work explores the application of multiscale stochastic averaging methods developed by~\citet{Majda2001cpam},~\citet{Givon2004nonlinearity, EWeinan2011book, Pradas2012},~\citet{PavliotisStuart2012}, and others to the closure problem in reduced-order models of incompressible flow or other systems reducible to linear-quadratic dynamics~\citep{Rowley2004pd, Qian2020, Kaptanoglu2021pre}, including introducing an approximation to the form of the fast dynamics that allows for computation of the averaged dynamics in closed form.
The procedure is summarized in Fig.~\ref{fig: intro -- overview} and shown schematically in Fig.~\ref{fig: intro -- nonlinearity}.
We refer to this secondary dimensionality reduction of the Galerkin system as multiscale model reduction (MMR).

Since fluid flows generally do not have a true scale separation away from the threshold of instability, we demonstrate via numerical simulations that this method is a robust and systematic approach to stabilizing low-dimensional models.
This extends the work of~\citet{Majda2006nonlinearity} exploring the application of this class of methods to systems beyond the parameter regimes where their validity can be rigorously proven.
Mltiscale model reduction is a unified framework for understanding the origin and importance of cubic terms in reduced-order models of the linear-quadratic Navier-Stokes equations, capable of capturing both mean flow deformation and subscale dissipation.
Throughout this work we also highlight connections to other modeling methods, including Koopman theory and weakly nonlinear analysis.


%% file: S2_Galerkin.tex
\section{Reduced-order modeling for incompressible flows}
\label{sec: galerkin}

For more than three decades, proper orthogonal decomposition (POD) and Galerkin projection have been the foundational tools in low-order modeling for nonlinear incompressible fluid flow.
POD analysis identifies dominant coherent structures in the flow and provides an energy-optimal linear modal basis. 
Galerkin projection approximates both the state of the flow and its time derivative in this finite-dimensional subspace, resulting in a reduced system of ordinary differential equations in terms of the POD mode amplitudes. 
These topics have been covered extensively elsewhere (see e.g.~\citet{Holmes1996, Noack2011book,Benner2015siam, Rowley2017, Taira2017}); here we provide a brief overview. 

We assume the flow is governed by the unsteady incompressible Navier-Stokes equations
  \begin{equation}
    \begin{aligned}
      \displaystyle \frac{\partial \bm{u}}{\partial t} + \nabla \cdot \left( \bm{u} \otimes \bm{u} \right) & = - \nabla p + \frac{1}{\Re} \nabla^2 \bm{u} \\
      \nabla \cdot \bm{u} & = 0,
    \end{aligned}
    \label{eq: navier-stokes}
  \end{equation}
where $\bm{u}(\bm{x}, t)$ is the velocity field, $p$ is the pressure field, and $\Re$ is the Reynolds number, $\Re=UL\nu$, based on the kinematic viscosity $\nu$ and suitable length and velocity scales $L$ and $U$.
The nonlinear term is expressed in divergence form with the tensor product $\otimes$.
In this work, we only consider two dimensional examples for which $\bm u = \begin{bmatrix} u & v \end{bmatrix}^T$, although all of the following applies equally well to three-dimensional flows.

More broadly, POD can be used for any collection of data for which a suitable norm can be defined.
The specific method described here for reducing the Navier-Stokes equations to a linear-quadratic system of ODEs via Galerkin projection relies on the quadratic nonlinearity in Eq.~\eqref{eq: navier-stokes}, although with certain restrictions this procedure can also be applied to systems such as compressible flow~\citep{Rowley2004pd}, magnetohydrodynamics~\citep{Kaptanoglu2021pre}, or systems with non-polynomial nonlinearities that have been ``lifted'' to a coordinate system in which the dynamics are linear-quadratic~\citep{Qian2020}.

\subsection{Proper orthogonal decomposition}
\label{sec: galerkin -- pod}

In order to reduce the system of partial differential equations~\eqref{eq: navier-stokes} to a finite-dimensional system of ordinary differential equations, we assume that the space and time dependence can be separated, so that an arbitrary velocity field $\bm u(\bm x, t)$ can be approximated with the $r$-dimensional modal representation
\begin{equation}
    \label{eq: modal-ansatz}
    \bm u(\bm x, t) \approx \bm{\psi}_0(\bm x) + \sum_{i=1}^r \bm \psi_i(\bm x) a_i(t),
\end{equation}
where $\bm{\psi}_0(\bm x)$ is a fixed base flow around which the expansion is performed and each mode individually satisfies the divergence-free constraint $\nabla \cdot \bm{\psi}_i = 0$ for $i=0, 2, \dots, r$.
Typically in this work we will take $\bm{\psi}_0$ to be the mean flow $\bar{\bm u}$ as estimated with a time average, although another common choice is to let $\bm{\psi}_0$ be the steady-state solution to the Navier-Stokes equations.

The proper orthogonal decomposition (POD) identifies the set of modes $\{\bm{\psi}_i\}_{i=1}^r$ that provide the optimal rank-$r$ approximation for an average velocity field in the energy norm induced by the velocity inner product
\begin{equation}
\label{eq: inner-product}
    \langle \bm u, \bm u' \rangle = \int_\Omega \bm u'(\bm x) \cdot \bm u^*(\bm x) \dd\, \Omega
\end{equation}
defined on the spatial domain $\Omega$, where the star indicates complex conjugation.
Since our data may be represented on a nonuniform mesh, we approximate this integral with a weighted sum
\begin{equation}
\label{eq: galerkin -- weighted-inner-product}
    \langle \bm u, \bm v \rangle \approx \mathsfbi{u}^T \mathsfbi{W} \mathsfbi{v},
\end{equation}
where $\mathsfbi{u}$ and $\mathsfbi{v}$ are the discrete approximations to $\bm u$ and $\bm v$, and $\mathsfbi{W}$ is a diagonal weight matrix containing the cell volumes.
The transpose in Eq.~\eqref{eq: galerkin -- weighted-inner-product} is understood to include complex conjugation in the case that the fields or modes are complex valued.

The POD modes themselves are eigenfunctions of the two-point spatial correlation tensor, although in practice they are typically approximated numerically with the method of snapshots~\citep{Sirovich1987} or the singular value decomposition~\citep{Brunton2019book}.
In the method of snapshots, the POD modes are estimated from a weighted sum of velocity field snapshots, where the weights are determined from eigenvectors of the two-point temporal correlation matrix.
The construction from a linear combination of divergence-free snapshots also ensures that the velocity modes are divergence-free.

The POD expansion~\eqref{eq: modal-ansatz} is often only applied to the velocity fields.
This can pose a problem for model reduction methods approximating the pressure gradient term in the Navier-Stokes equations.
While this term vanishes for closed flows and is often negligible for flows with a localized global instability, it is important in open shear flows, such as the mixing layer examined in Sec.~\ref{sec: results}~\citep{Noack2005jfm}.
There are various methods for approximating the pressure term (see e.g.~\citet{Caiazzo2014jcp}), but in this work we use a velocity-pressure expansion with the same inner product~\eqref{eq: inner-product}.
Defining $\bm q(\bm x, t) = \begin{bmatrix} \bm u & p \end{bmatrix}^T$, we replace~\eqref{eq: modal-ansatz} and~\eqref{eq: inner-product} with
\begin{subequations}\label{eq: galerkin -- velocity-pressure-pod}
\begin{align}
    \bm q(\bm x, t) &\approx
    \begin{bmatrix}
     \bm \psi^{\bm u}_0(\bm x) \\ \psi^p_0(\bm x)
    \end{bmatrix} + \sum_{i=1}^r \begin{bmatrix}
     \bm \psi^{\bm u}_i(\bm x) \\ \psi^p_i(\bm x)
    \end{bmatrix} a_i(t) \\
    &= \bm{\psi}_0(\bm x) + \sum_{i=1}^r \bm \psi_i(\bm x) a_i(t)
\end{align}
\end{subequations}
and energy inner product $\langle \bm q, \bm q' \rangle_{\bm u} \equiv \langle \bm u, \bm u' \rangle$.
Numerically, we simply carry the pressure fields through the method of snapshots by setting the diagonal entries of $\mathsfbi{W}$ corresponding to pressure to zero.
The pressure components of the POD modes are therefore estimated from the same linear combination of snapshots as the velocity components.

Besides providing an optimal linear modal approximation, the POD representation has the advantage that the velocity modes are orthonormal and the temporal coefficients are uncorrelated in time:
\begin{subequations}
\label{eq: pod-properties}
\begin{gather}
    \langle \bm \psi_i, \bm \psi_j \rangle_{\bm u} = \delta_{ij} \\
    \overline{a_i a_j} = \lambda_i \delta_{ij},
\end{gather}
\end{subequations}
where the constant of proportionality $\lambda_i$ for $\overline{a_i^2}$ is the POD eigenvalue, with the physical meaning of average fluctuation kinetic energy in the $i^\mathrm{th}$ mode.
The subspace dimension $r$ is typically selected based on residual energy content of the truncated modes.
Once the modes $\{ \bm \psi_i \}_{i=1}^r$ are calculated, the temporal coefficients $\bm a(t)$ can be computed by projecting the time series onto the span of the modes, i.e. $a_i(t) = \langle \bm \psi_i(\bm x), \bm u(\bm x, t) \rangle $.

Although POD analysis is a powerful tool, it suffers from several limitations.
First, it assumes a space-time separation of variables, which is typically appropriate for closed flows, but is often a poor approximation of inhomogeneous open flows dominated by traveling waves.
Second, the condition that the temporal coefficients be linearly uncorrelated does not rule out nonlinear correlations.
Finally, the optimality guarantee of the POD approximation is purely based on the instantaneous flow field and does not account for any time dynamics.
The relationship between the first two issues was examined at length by~\citet{Callaham2022nonlinear}, where it was shown that leveraging nonlinear correlations can stabilize reduced-order models of advection-dominated flows while reducing the dimensionality of the model. 
The third issue motivates alternative modal representations, such as dynamic mode decomposition~\citep{Rowley2009, Schmid2010,Kutz2016book,schmid2022dynamic}, spectral POD~\citep{Towne2018jfm,schmidt2020guide}, balanced POD~\cite{Willcox2002aiaaj,Rowley2005ijbc}, resolvent modes~\citep{McKeon2010jfm,Sharma2013jfm,zare2017colour,Padovan2020jfm}, exact coherent structures~\citep{graham2020exact,Sharma2016prf,rosenberg2019efficient}, and trajectory-based optimization~\citep{Otto2022}. 
These drawbacks notwithstanding, we use POD in this work due to its numerical robustness and orthonormality properties.

\subsection{Galerkin projection}
Galerkin projection is a widely used method in computational and applied mathematics for discretizing continuous problems, in particular partial differential equations (PDEs). 
In the context of reduced-order modeling, this approach typically refers to projecting the infinite-dimensional PDE onto a finite-dimensional subspace spanned by global, orthogonal basis functions.
While such a basis can by constructed analytically for simple geometries, the energy optimality and orthonormality of POD makes it a natural candidate for statistically stationary (i.e. post-transient) flows with more complex geometries.  
Moreover, it is a data-driven approach that tailors the basis based on observed flow kinematics.  

The derivation of the Galerkin system begins by substituting the velocity-pressure POD approximation~\eqref{eq: galerkin -- velocity-pressure-pod} into the Navier-Stokes equations~\eqref{eq: navier-stokes}.
In order for the dynamics to be an optimal continuous-time approximation, the residual error should be orthogonal to the POD subspace.
Using the inner product~\eqref{eq: inner-product} and orthonormality properties~\eqref{eq: pod-properties}, this optimality condition leads to the linear-quadratic system of ODEs~\citep{Holmes1996, Noack2011book}
\begin{equation}
    \label{eq: galerkin -- linear-quadratic}
    \dot{a}_i = F_i + L_{ij} a_j + Q_{ijk} a_j a_k, \qquad i, j, k = 1, 2, \dots, r
\end{equation}
with constant, linear, and quadratic terms given by
\begin{subequations}
\begin{align}
    F_i &= \langle \bm \psi_i, -\nabla \cdot (\bm \psi_0^{\bm u} \otimes \bm \psi_0^{\bm u} ) - \nabla \psi_0^{p} + \Rey^{-1} \nabla^2 \bm \psi_0^{\bm u} \rangle_{\bm u} \\
    L_{ij} &= \langle \bm \psi_i, -\nabla \cdot (\bm \psi_0^{\bm u} \otimes \bm \psi^{\bm u}_j +  \bm \psi^{\bm u}_j \otimes \bm \psi_0^{\bm u} ) - \nabla \psi^p_j + \Rey^{-1} \nabla^2 \bm \psi^{\bm u}_j \rangle_{\bm u} \\
    Q_{ijk} &= \langle \bm \psi_i,  -\nabla \cdot (\bm \psi_j^{\bm u} \otimes \bm \psi_k^{\bm u} ) \rangle_{\bm u}.
\end{align}
\end{subequations}
Throughout most of this work, we take the base mode to be the time-averaged mean flow.
However, if the flow is instead expanded about the steady-state solution, the constant forcing term $\mathsfbi{F}$ vanishes, ensuring that the origin is a fixed point.

\subsection{Symmetries and energy conservation in the Galerkin system}
\label{sec: galerkin -- symmetries}

The combination of proper orthogonal decomposition and Galerkin projection is a popular choice for model reduction of incompressible flows, in part because it provides a semi-empirical ``gray-box'' framework that combines some of the advantages of data-driven analysis with physics-based modeling.  
In particular, the modal basis is tailored from data for a particular flow configuration, while the dynamics are obtained by directly projecting the governing equations onto this basis.  
One consequence is that the dynamical system Eq.~\eqref{eq: galerkin -- linear-quadratic} inherits certain properties from the Navier-Stokes equations.
For example, although we will mostly take the base flow $\bm \psi_0(\bm x)$ to be the time-averaged mean flow, if the the flow is instead expanded about the steady-state solution the constant forcing term $\mathsfbi{F}$ vanishes, ensuring that the origin is a fixed point.
If, in addition, an energy inner product such as Eq.~\eqref{eq: inner-product} is used for projection, the reduced-order model also preserves Lyapunov stability of the origin, although not necessarily the stability of limit cycles or other fixed points~\citep{Rowley2004pd}.

A key feature of the quadratic nonlinearity of the Navier-Stokes equations is that it is energy-preserving, in the sense that it has no net contribution to the evolution equation for kinetic energy in the spectral domain~\citep{Kraichnan1989}, with some restrictions on the boundary conditions.
This is the foundation of the energy cascade picture of turbulence, in which the role of the nonlinearity is to transfer energy from the large scales to the small, dissipative scales. 
For the Galerkin system, the energy preservation condition implies that~\citep{Schlegel2015jfm}
\begin{equation}
    Q_{ijk} + Q_{ikj} + Q_{jik} + Q_{jki} + Q_{kij} + Q_{kji} = 0.
\end{equation}
This is often approximately true numerically for POD-Galerkin systems, but enforcing it explicitly can improve the stability of the model~\citep{Cordier2013}.
Based on the structure of the linear-quadratic system~\eqref{eq: galerkin -- linear-quadratic}, the quadratic tensor can also be symmetrized in the last two indices:
\begin{equation}
    Q_{ijk} = Q_{ikj}.
\end{equation}

Finally,~\citet{Schlegel2015jfm} showed that the long-term boundedness of the Galerkin system can be guaranteed if a criterion based on generalized Lyapunov functions is satisfied.
Although this is a dynamical systems analysis and not an \emph{a priori} theoretical guarantee, it can also be used as a constraint in sparse model identification to construct empirical Galerkin-type models with guaranteed stability~\citep{Kaptanoglu2021}.

%% file: S3_Generators.tex
\section{Generators of deterministic and stochastic processes}
\label{sec: generators}

In the study of mechanics there are often several equivalent representations of the same system (e.g. Lagrangian, Hamiltonian, etc.), each of which may be useful depending on the application.
The operator-theoretic perspective will be especially useful in the development of the multiscale closure model as a framework for abstract formal manipulations of the dynamical systems model.
In particular, the closure model will be derived via a perturbation series approximation to the action of a stochastic Koopman operator.
Critically, the infinite-dimensional operator itself need not be computed or directly approximated; instead, the closed model appears as a solvability condition analogous to the derivation of the amplitude equation in weakly nonlinear analysis~\citep{Sipp2007jfm}.

This section gives a brief overview of topics necessary for the development of the closure models in Section~\ref{sec: closure}; for a more in-depth presentation of these topics see, e.g.~\citet{Risken1996book, EWeinan2011book, PavliotisStuart2012, klus2018data,Klus2020physica, Brunton2022koopman}.
Koopman theory is perhaps the most widely-discussed operator-theoretic method in recent work on the analysis of fluid flows and other large-scale nonlinear dynamics~\citep{Mezic2005,Rowley2009,Mezic2013arfm,klus2018data,Brunton2022koopman}, making it a convenient place to begin the discussion.

Suppose the state $\bm{x}(t)$ is governed by an autonomous ordinary differential equation (ODE)
\begin{equation}
\label{eq: generators -- ode}
    \dot{\bm{x}} = \bm{f}(\bm{x}), \qquad \bm x \in \mathcal{X}.
\end{equation}
We will assume, unless otherwise specified, that states are real-valued,  e.g. $\mathcal{X} \subseteq \R^N$, and that all other functions are $L^2$-integrable with inner product
\begin{equation}
    \langle f(\bm x), g(\bm x) \rangle = \int_\mathcal{X} f(\bm x) g(\bm x) \dd x,
\end{equation}
where the usual complex-conjugation is omitted since $g$ is real-valued.

The Koopman operator $\mathcal{K}^t$ acts on the $L^2$-integrable space of scalar observables $g(\bm{x}): \mathcal{X} \rightarrow \R$, advancing them forward time $t$.
More precisely, let $\bm x$ be the solution to the initial value problem of~\eqref{eq: generators -- ode} with $\bm x(0) = \bm x_0$. 
Then the Koopman operator is defined as
\begin{equation}
    \mathcal{K}^t g(\bm{x}_0) = g(\bm x(t)). 
\end{equation}

Although the Koopman operator is generally difficult to either represent explicitly or approximate in a useful finite-dimensional subspace, analysis of its spectral properties has drawn great interest in recent work.
For our purposes, the infinitesimal generator $\mathcal{L}$, defined by
\begin{equation}
    \mathcal{L}g = \lim_{t \rightarrow 0} \frac{\mathcal{K}^t g - g}{t}
\end{equation} and sometimes called the Lie operator, is more theoretically useful.
If we consider $g(\bm x(t))$ to be an explicit function of time, then $\mathcal{L}$ can be derived by applying the chain rule to $g$.
With a slight abuse of notation, if $g(\bm x, t)$ is taken to be a function of both time and initial state $\bm x$ (i.e. in the Lagrangian frame of reference), then this can be extended to a PDE over all of state space:
\begin{equation}
\label{eq: generators -- koopman-generator}
    \pdv{g}{t} = \mathcal{L} g = \bm f(\bm x) \cdot \nabla g.
\end{equation}
Thus the generator $\mathcal{L}$ is a linear advection operator governing the evolution of scalar observables of the system described by~\eqref{eq: generators -- ode}.
Similarly, the adjoint of \eqref{eq: generators -- koopman-generator},
\begin{equation}
\label{eq: generators -- perron-frobenius-generator}
    \pdv{\rho}{t} = \mathcal{L}^\dagger \rho = - \nabla \cdot (\rho \bm f(\bm x)),
\end{equation}
is a continuity equation governing the evolution of densities $\rho(\bm x, t)$ in phase space.
As a point of reference, equation~\eqref{eq: generators -- perron-frobenius-generator} reduces to the Liouville equation from classical mechanics under the incompressibility condition $\nabla \cdot \bm f = 0$, and $\mathcal{L}^\dagger$ is also known as the generator of the Perron-Frobenius operator~\citep{Froyland2005physd,Froyland2009pd,froyland:2010,klus2016numerical}.

This description of dynamics can readily be extended to systems governed by stochastic differential equations (SDEs) of the form
\begin{equation}
\label{eq: generators -- sde}
    \dot{\bm x} = \bm f(\bm x) + \bm \Sigma \bm w(t),
\end{equation}
where the deterministic component $\bm f(\bm x)$ is known as the drift function and the diffusion matrix $\bm \Sigma$ modifies a vector-valued Gaussian white noise process $\bm w(t)$.
In this case the evolution of the probabilitiy distribution $\rho(\bm x, t)$ is governed by the stochastic analog of~\eqref{eq: generators -- perron-frobenius-generator}, known as the forward Kolmogorov, or Fokker-Planck, equation:
\begin{equation}
\label{eq: generators -- fokker-planck}
    \pdv{\rho}{t} = \mathcal{L}^\dagger \rho = -\nabla \cdot (\rho \bm f) + \nabla \nabla^T : (\rho \bm D),
\end{equation}
where $\bm D = \bm \Sigma \bm \Sigma^T / 2$ is the diffusion tensor and the colon denotes tensor contraction.
The associated backwards Kolmogorov equation is the adjoint of \eqref{eq: generators -- fokker-planck}:
\begin{equation}
\label{eq: generators -- bke}
    \pdv{g}{t} = \mathcal{L} g = \bm f \cdot \nabla g + \bm D : \nabla \nabla^T g.
\end{equation}

To interpret \eqref{eq: generators -- bke}, consider the expectation of a scalar observable $g(\bm x): \mathcal{X} \rightarrow \R$ defined by the inner product of $g$ with the probability distribution $\rho_0(\bm x)$:
\begin{equation}
    \E[g] = \langle \rho_0, g \rangle = \int_\mathcal{X} \rho_0(\bm x) g(\bm x) \dd \bm x.
\end{equation}
The probability distribution is advanced in time by $\rho(\bm x, t) = e^{\mathcal{L}^\dagger t} \rho_0(\bm x)$.
Using the definition of the adjoint, $\langle \rho, \mathcal{L} g \rangle = \langle \mathcal{L}^\dagger \rho, g \rangle$, the expectation of $g$ at time $t$ can be written equivalently as
\begin{equation}
    \E[g](t) = \langle \rho, g \rangle = \int_\mathcal{X} \rho_0(\bm x) e^{\mathcal{L} t} g(\bm x) \dd \bm x.
\end{equation}
Therefore, just as the Koopman operator advances an observable in time, the backwards Kolmogorov equation describes the evolution of the expectation of an observable.

As a simple example relevant for the closure modeling presented below, consider the Ornstein-Uhlenbeck process defined by the SDE
\begin{equation}
    \dot{x} = \nu(\mu - x) + \sigma w(t)
\end{equation}
for positive constants $\nu$, $\mu$, and $\sigma$.
The stationary distribution $\rho^\infty(x)$ is given by the steady-state Fokker-Planck equation
\begin{equation}
\label{eq: generators -- OU-process}
    0 = \mathcal{L}^\dagger \rho^\infty = -\dv{x} \rho^\infty(x) \nu (\mu - x) + \frac{\sigma^2}{2}\dv[2]{x} \rho^\infty(x),
\end{equation}
along with the usual normalization condition on $\rho^\infty$.
This can be solved analytically by a Gaussian distribution with mean $\mu$ and variance $\sigma^2/2\nu$.
Defining an observable that is the state itself $g(x) = x$, the evolution is given by the backwards Kolmogorov equation~\eqref{eq: generators -- bke}, which in this case reduces to the initial value problem
\begin{equation}
    \dv{g}{t} = \nu(\mu - g), \qquad g(0) = x.
\end{equation}
The expectation at time $t$ is the solution $g(t)$, which in this case is exponential decay towards the mean $\mu$:
\begin{equation}
    \E[x](t) = g(t) = \mu - (x - \mu) e^{-\nu t}.
\end{equation}

Given that the model reduction procedure described in Section~\ref{sec: galerkin} aims to reduce the physics model from an infinite-dimensional partial differential equation to a finite-dimensional system of ordinary differential equations, it may be counterintuitive that it would be helpful to return to an infinite-dimensional function space and represent the dynamics with a partial differential equation.
The primary advantage in doing so is that these generators are linear operators, which in some cases can be amenable to approaches that are unavailable for the nonlinear dynamics of the ODE~\eqref{eq: generators -- ode} or SDE~\eqref{eq: generators -- sde}.

Aside from simple one-dimensional or linear examples like the Ornstein-Uhlenbeck process, it is typically very difficult to solve these PDEs analytically, motivating the approximation in terms of an asymptotic expansion in Sec.~\ref{sec: closure}.
As will be seen, this expansion can be used to derive a solvability condition based on the Fredholm alternative.
Using the manipulations reviewed in this chapter, particularly in terms of the generators and their adjoints, this condition can be interpreted as an average over subscale variables, which can be computed explicitly under suitable assumptions.
The averaged slow dynamics then take the form of a cubic generalized Stuart-Landau equation that accounts for the leading-order effect of the unresolved fast variables.

%% file: S4_Closure.tex
\section{Multiscale closure modeling}
\label{sec: closure}

One of the primary difficulties of reduced-order modeling for fluid flows is that the nonlinear interactions in the Navier-Stokes equations transfer energy between all scales of the flow. 
Thus, restricting the dynamics to a low-dimensional subspace can lead to significant approximation errors. 
Generally speaking, the goal of closure modeling is to augment this truncated model with additional terms that account for the effect of the unresolved scales. 

The multiscale approach to closure modeling accomplishes this by first partitioning the dynamics into fast and slow variables, and then approximating the solution to the associated backwards Kolmogorov equation with a perturbation series expansion.
The result can be interpreted as a Koopman generator of the form of Eq.~\eqref{eq: generators -- koopman-generator} corresponding to the coarse-grained dynamics for the slow variables.
As we will show, when applied to the linear-quadratic Galerkin system~\eqref{eq: galerkin -- linear-quadratic}, this leads to a generalized Stuart-Landau equation including cubic terms.

An interesting feature of cubic Stuart-Landau-type models of fluid flow is that they are often more accurate than linear-quadratic models, even though the nonlinearity in the underlying governing equations is quadratic~\citep{Noack2003jfm, Loiseau2017jfm}.
One reason for this is the artificial space-time separation of variables introduced by the modal representation~\eqref{eq: modal-ansatz}, which can introduce spurious degrees of freedom that represent phase-locked harmonics of traveling waves, for instance~\citep{Callaham2022nonlinear}.
As a result, POD-Galerkin models are vulnerable to decoherence and can fail to improve with increasing rank, even when the kinematic approximation of the flow field becomes nearly perfect.
As illustrated by the examples in Sec.~\ref{sec: results}, the introduction of cubic terms can mitigate this, either by modifying the dynamics to resemble phase-locked nonlinear oscillators or by eliminating spurious degrees of freedom altogether.

Although multiscale methods can be rigorously justified when there is a strict separation of timescales~\citep{Majda2001cpam, EWeinan2003, EWeinan2011book, PavliotisStuart2012}, there is no such spectral gap in most fluid flows of practical interest.
The proposed method for model reduction should therefore properly be viewed as an approximate closure model motivated by the asymptotic limit in the spirit of the variational multiscale approach to turbulence modeling~\citep{Bazilevs2007}.
Nevertheless, it provides a systematic method for stabilizing low-dimensional models and also highlights a generic mechanism by which higher-order nonlinearities can arise from the quadratic term in the governing equations.

\subsection{Averaging over unresolved variables}
\label{sec: closure -- averaging}

Beginning with the Galerkin dynamics~\eqref{eq: galerkin -- linear-quadratic} for the state consisting of POD coefficients $\bm{a}(t) \in \R^r$, where we assume that $r$ is large enough for an accurate kinematic reconstruction of a typical flow field, we partition the system into slow variables $\bm x(t) \in \mathcal{X} = \R^{r_0}$ and fast variables $\bm y(t) \in \mathcal{Y} = \R^{r-r_0}$, so that $\bm a = \begin{bmatrix} \bm x^T & \bm y^T \end{bmatrix}^T$.

For concise notation we will use the Einstein convention that repeated indices imply summation and we will omit explicit summation unless not doing so would lead to ambiguity.
We will index the slow variables with Roman subscripts $i$, $j$, $k$, $\ell$ ranging from $1$ to $r_0$ and the fast variables with Greek subscripts $\alpha$, $\beta$, $\gamma$ ranging from $1$ to $r-r_0$.
Without loss of generality we will also assume that the quadratic term has been symmetrized in the last two indices, so that $Q_{ijk} = Q_{ikj}$.
Then the partitioned Galerkin system is
\begin{subequations}
\label{eq: closure -- partitioned-galerkin}
\begin{align}
    \dot{x}_i &= f^x_i(\bm x, \bm y) = F^1_i + L^{11}_{ij} x_j + L^{12}_{i\beta} y_\beta + Q^{111}_{ijk} x_j x_k + 2 Q^{112}_{ij\beta} x_j y_\beta + Q^{122}_{i \beta \gamma} y_\beta y_\gamma
    \label{eq: closure -- partitioned-galerkin:x}\\ 
    \dot{y}_\alpha &= f^y_\alpha(\bm x, \bm y) = F^2_\alpha + L^{21}_{\alpha j} x_j + L^{22}_{\alpha\beta} y_\beta + Q^{211}_{\alpha jk} x_j x_k + 2 Q^{212}_{\alpha j \beta} x_j y_\beta + Q^{222}_{\alpha \beta \gamma} y_\beta y_\gamma.
    \label{eq: closure -- partitioned-galerkin:y}
\end{align}
\end{subequations}

Standard truncation of this system is equivalent to retaining only the terms $\mathsfbi{F}^1$, $\mathsfbi{L}^{11}$, and $\mathsfbi{Q}^{111}$.
Here we will attempt to approximate the terms involving the fast variables in $\bm f^x(\bm x, \bm y)$ in an average sense in order to derive a closed system $\dot{\bm x} = \hat{\bm f}(\bm x)$, largely following the approach of~\citet{PavliotisStuart2012}.

We assume that the time scales of the fast and slow dynamics are separated by a parameter $\epsilon \ll 1$, so that we may define $\bm f^x(\bm x, \bm y) \equiv \epsilon \tilde{\bm f}^x(\bm x, \bm y)$.
Furthermore, we note that the role of the fast self-interaction term $\mathsfbi{Q}^{222}(\bm y, \bm y)$ is to transfer energy between the unresolved scales.
Since this mechanism is of secondary importance to the transfers between slow and fast scales, we apply a version of the ``working assumption of stochastic modeling''~\citep{Majda2001cpam}:
\begin{equation}
    Q^{222}_{\alpha \beta \gamma} y_\beta y_\gamma \approx \sigma_{\alpha} w_\alpha(t),
\end{equation}
where $\bm w(t)$ is a Wiener process and $\bm \sigma$ is an as-yet-undefined constant forcing amplitude.
Finally, we coarse-grain the dynamics on the time scale $\tau = \epsilon t$, leading to the slow/fast system
\begin{subequations}
\label{eq: closure -- slow-fast-galerkin}
\begin{align}
    \pdv{x_i}{\tau} &= \tilde{f}^x_i(\bm x, \bm y) \\ 
    \pdv{y_\alpha}{\tau} &= \frac{1}{\epsilon} f^y_\alpha(\bm x, \bm y) +  \frac{1}{\sqrt{\epsilon}} \sigma_{\alpha} w_\alpha (\tau),
\end{align}
\end{subequations}
where we have also used the scaling property of Wiener processes that $\bm w(\epsilon t) = \sqrt{\epsilon} \bm w(t)$.

Defining an arbitrary scalar-valued observable $g^\epsilon(\bm x, \bm y)$, the backwards Kolmogorov equation~\eqref{eq: generators -- bke} associated with Eq.~\eqref{eq: closure -- slow-fast-galerkin} is
\begin{equation}
\label{eq: closure -- bke}
    \pdv{g^\epsilon}{\tau} = \frac{1}{\epsilon} \underbrace{\left[ f^y_\alpha(\bm x, \bm y) \pdv{y_\alpha} + \frac{\sigma_\alpha^2}{2} \pdv[2]{y_\alpha}  \right]}_{\mathcal{L}_0} g^\epsilon  + \underbrace{ \tilde{f}^x_i(\bm x, \bm y) \pdv{x_i} }_{\mathcal{L}_1} g^\epsilon.
\end{equation}
Note that $\mathcal{L}_0$ can be viewed as the generator of a stochastic process in $\bm y$ with $\bm x$ as a fixed parameter.
Consequently, we make the following assumptions related to the ergodicity of $\bm y$:
\begin{enumerate}
    \item The operator $\mathcal{L}_0$ has a one-dimensional nullspace spanned by constants in $\bm y$:
    \begin{equation}
    \label{eq: closure -- nullspace-fwd}
        \mathcal{L}_0 g(\bm x) = 0.
    \end{equation}
    \item The Fokker-Planck operator $\mathcal{L}_0^\dagger$ has a one-dimensional nullspace corresponding to the stationary distribution $\rho^\infty_{\bm x}$, where again $\bm x$ is treated as a fixed parameter:
    \begin{equation}
    \label{eq: closure -- nullspace-adj}
        \mathcal{L}_0^\dagger \rho^\infty_{\bm x}(\bm y) = 0,
    \end{equation}
    along with the usual normalization condition $ \int_\mathcal{Y} \rho^\infty_{\bm x} \dd \bm y = 1 $.
\end{enumerate}
These assumptions obviate the need to express the fast scale $\bm y$ as an instantaneous function of $\bm x$, as in an invariant manifold model~\citep{GuckenheimerHolmes, PavliotisStuart2012}.
Instead, the fast variable is modeled with a simplified distribution that can be averaged over as follows.
We assume that the backwards Kolmogorov equation~\eqref{eq: closure -- bke} can be approximated by means of an asymptotic expansion
\begin{equation}
\label{eq: closure -- perturbation-series}
    g^\epsilon(\bm x, \bm y, \tau) = g_0 + \epsilon g_1 + \mathcal{O}(\epsilon^2).
\end{equation}
Substituting into Eq.~\eqref{eq: closure -- bke} and equating powers of $\epsilon$ gives the consistency conditions
\begin{subequations}
\begin{align}
\label{eq: closure -- consistency-order-0}
    0 &= \mathcal{L}_0 g_0, \\
\label{eq: closure -- consistency-order-1}
    \pdv{g_0}{\tau} &= \mathcal{L}_0 g_1 + \mathcal{L}_1 g_0.
\end{align}
\end{subequations}
By virtue of the assumption of a one-dimensional nullspace, Eq.~\eqref{eq: closure -- consistency-order-0} is satisfied if $g_0$ is not a function of $\bm y$, or $g_0 = g_0(\bm x, \tau)$.
As expected, the leading-order solution does not depend on the fast variable.

An effective evolution equation for $g_0$ can be derived by considering Eq.~\eqref{eq: closure -- consistency-order-1} as a linear equation for $g_1(\bm x, \bm y, \tau)$.
The Fredholm alternative specifies that for any equation of the form
$
    \mathcal{L}_0 g_1 = b
$
to have a unique solution, all functions $\rho$ in the nullspace of the adjoint operator $\mathcal{L}_0^\dagger$ must be orthogonal to $b$.
Since we have assumed that the nullspace of $\mathcal{L}_0^\dagger$ is one-dimensional and spanned by the stationary distribution $\rho_{\bm x}^\infty(\bm y)$, this implies that $\langle \rho_{\bm x}^\infty, b \rangle = 0$.
In other words, the solvability condition for
\begin{equation}
\label{eq: closure -- fredholm-equation}
    \mathcal{L}_0 g_1 = -\pdv{g_0}{\tau} + \mathcal{L}_1 g_0
\end{equation}
is that the right-hand side of Eq.~\eqref{eq: closure -- fredholm-equation} has zero mean with respect to the stochastic process generated by $\mathcal{L}_0$:
\begin{equation}
\label{eq: closure -- fredholm-alternative}
    \int_{\mathcal{Y}} \rho_{\bm x}^\infty(\bm y) \left[ -\pdv{\tau} + \mathcal{L}_1 \right] g_0(\bm x, \tau) \dd \bm y = 0.
\end{equation}
Using the normalization condition for $\rho_x^\infty$,  this simplifies to a closed evolution equation for $g_0(\bm x, \tau)$:
\begin{equation}
\label{eq: closure -- closed-bke}
    \pdv{g_0}{\tau} = \left[ \int_{\mathcal{Y}} \rho_{\bm x}^\infty(\bm y) \tilde{f}_i^x(\bm x, \bm y) \dd \bm y \right] \pdv{g_0}{x_i}.
\end{equation}
Again, the observable $g_0$ itself is arbitrary, but Eq.~\eqref{eq: closure -- closed-bke} has the form of the generator of a Koopman operator corresponding to the coarse-grained dynamics in $\bm x$ alone.
Undoing the $\epsilon$ scaling in $\tau$ and $\tilde{\bm f}^x$, Eq.~\eqref{eq: closure -- closed-bke} can be written as
\begin{equation}
\label{eq: closure -- closed-bke-rescaled}
\pdv{g_0}{t} = \hat{\bm f} (\bm x) \cdot \nabla g_0,
\end{equation}
corresponding to the averaged dynamics
\begin{subequations}
\label{eq: closure -- closed-ode}
\begin{align}
    \dot{\bm x} &= \hat{\bm f}(\bm x), \\
    \label{eq: closure -- closed-ode-avg}
    \hat{\bm f}(\bm x) &= \int_{\mathcal{Y}} \rho_{\bm x}^\infty(\bm y) f_i^x(\bm x, \bm y) \dd \bm y.
\end{align}
\end{subequations}

The distribution $\rho_{\bm x}^\infty(\bm y)$ specifies the probability distribution of the fast variables $\bm y$ and is stationary on the fast timescale, but implicitly time-varying since it is parameterized by the state $\bm x$ of the slow variables. 
In the simplest case $\rho_{\bm x}^\infty$ might be proportional to a delta function in $\bm x$, indicating that the fast variables are a direct function of the slow variables.
Physically this might correspond to the case where $\bm y$ represents a slow amplitude-dependent deformation of the base flow or phase-locked higher harmonics, for instance.

More generally the functional form of this distribution might be complicated, but it is not necessary to specify it in closed form provided the integral in Eq.~\ref{eq: closure -- closed-ode-avg} can be evaluated.
For instance, when the dynamics are linear-quadratic then Eq.~\ref{eq: closure -- closed-ode-avg} simplifies to first and second moments of the distribution.
Then the key to the MMR closure for the POD-Galerkin is deriving an approximation of the fast dynamics that is consistent with the original system but allows for computation of these moments in closed form.

As an aside, although the disappearance of the fictitious small parameter $\epsilon$ is necessary for consistency with the original Galerkin system, it does call into question the validity of the perturbation series approximation~\eqref{eq: closure -- perturbation-series}.
In this work we do not attempt to make this more rigorous, but instead demonstrate by example that it is a useful heuristic capable of resolving important features of the flow physics.
We justify this based on the observation that the underlying linear-quadratic Galerkin systems often poorly approximate the true dynamics, so it would not be useful to perfectly match the original model even if it was possible to do so.

\subsection{Stochastic averaging for a model system}
\label{sec: closure -- example}

Before applying the formal solution ~\eqref{eq: closure -- closed-ode} to the linear-quadratic Galerkin dynamics, we first illustrate the stochastic averaging procedure on a simple two-dimensional fast-slow system:
\begin{subequations}
\begin{align}
    \dv{x}{\tau} &= \lambda x - xy \\
    \epsilon \dv{y}{\tau}  &= -y + \mu x^2 + \sqrt{\epsilon} \sigma w(t).
\end{align}
\end{subequations}
This SDE system is associated with the following backwards Kolmogorov equation for an arbitrary scalar observable $g^\epsilon(x, y, \tau)$:
\begin{equation}
\label{eq: example -- bke}
    \pdv{g^\epsilon}{\tau} =
    \epsilon^{-1} \underbrace{ \left[ (-y + \mu x^2) \pdv{y} + \frac{\sigma^2}{2} \pdv[2]{y} \right] }_{\mathcal{L}_0} g^\epsilon +
    \underbrace{ \left[ (\lambda x - x y) \pdv{x} \right] }_{\mathcal{L}_1} g^\epsilon.
\end{equation}
Here $\mathcal{L}_0$ is the backwards Kolmogorov operator associated with an Ornstein-Uhlenbeck process in $y$, which we assume satisfies the ergodicity requirements~\eqref{eq: closure -- nullspace-fwd} and~\eqref{eq: closure -- nullspace-adj}.
By comparison of $\mathcal{L}^\dagger_0$ with the steady-state Fokker-Planck equation for an Ornstein-Uhlenbeck process~\eqref{eq: generators -- OU-process},
the stationary distribution $\rho^\infty_x$ is a Gaussian with mean and variance
\begin{equation}
\label{eq: closure -- example-moments}
    \int_{-\infty}^\infty y \rho^\infty_x(y) \dd y = \mu x^2, \qquad \qquad
    \int_{-\infty}^\infty y^2 \rho^\infty_x(y) \dd y = \frac{\sigma^2}{2}.
\end{equation}
In this case, the solvability condition~\eqref{eq: closure -- fredholm-alternative} reduces to
\begin{equation}
\label{eq: closure -- example-fredholm}
    \pdv{g_0}{\tau} = \left[ \int_{-\infty}^\infty \rho_x^\infty(y) (\lambda x - x y) \dd y \right] \pdv{g_0}{x} = 0.
\end{equation}
Comparing with Eq.~\eqref{eq: closure -- closed-bke}, this condition corresponds to a Koopman operator for the averaged dynamics.
Using the normalization condition on the probability distribution $\rho_x^\infty$ and the mean given by Eq.~\eqref{eq: closure -- example-moments}, the fast variable can be integrated analytically, giving the coarse-grained dynamics
\begin{align}
    \label{eq: closure -- example-closed-ode}
    \dv{x}{\tau} &= \int_{-\infty}^\infty \rho_x^\infty(y) (\lambda x - xy) \dd y \\
    &=\lambda x - \mu x^3.
\end{align}

Finally, we note that in this case the same solution could have been reached by neglecting the diffusion term in the $y$ dynamics from the beginning, as in an invariant manifold-type model~\citep{GuckenheimerHolmes, Noack2003jfm, PavliotisStuart2012}, since due to the form of the slow dynamics the effect of the noise disappears on averaging.
This would give the approximation $y = \mu x^2$, which could then be substituted into the $x$-dynamics to recover Eq.~\eqref{eq: closure -- example-closed-ode}.
The difference between that approach and the one described here lies in Eq.~\eqref{eq: closure -- example-fredholm}; rather than assuming the fast variables can be expressed as an instantaneous function of the slow variables, they are assumed to fluctuate and their effect is accounted for by averaging.
Later we will discuss some general conditions under which the two are equivalent, but in this example if there was a term proportional to $y^2$ in the $x$ dynamics, the solvability condition would include the second moment of $y$ and the diffusion $\sigma$ would appear in the coarse-grained dynamics.

\subsection{Application to the Galerkin system}
\label{sec: closure -- galerkin}

As illustrated by the previous example, in order to make practical use of the averaging procedure for the partitioned Galerkin system~\eqref{eq: closure -- partitioned-galerkin}, we must be able to perform the integral over the distribution $\rho_{\bm x}^\infty(\bm y)$ of the fast variables.
This distribution is the solution for fixed $\bm x$ to the steady-state Fokker-Planck equation
\begin{equation}
    \mathcal{L}^\dagger_0 \rho = -\pdv{(\rho f_\alpha^y )} {y_\alpha}  + \frac{\sigma_\alpha^2}{2} \pdv[2]{\rho}{y_\alpha} = 0,
\end{equation}
corresponding to the linear stochastic process
\begin{equation}
\label{eq: closure -- multivariate-OU}
    \dot{y}_\alpha = \left[ F^2_\alpha + L^{21}_{\alpha j} x_j + Q^{211}_{\alpha j k} x_j x_k \right] + \left[ L^{22}_{\alpha \beta} + 2Q^{212}_{\alpha j \beta} x_j \right] y_\beta + \sigma_\alpha w_\alpha(t).
\end{equation}
The stationary distribution is a multivariate Gaussian with mean and covariance that can be determined analytically from the solution of a Lyapunov equation~\citep{Risken1996book}, but this would need to be done at each value of $\bm x$.
Instead, we propose the diagonal drift approximation
\begin{equation}
\label{eq: closure -- diagonal-drift}
    \dot{y}_\alpha \approx \nu_\alpha(\bm x) \left(\mu_\alpha(\bm x) - y_\alpha \right) + \sigma_\alpha w_\alpha (t),
\end{equation}
for which the stationary distribution is a product of univariate Gaussians solving the 1-dimensional Fokker-Planck equation~\eqref{eq: generators -- OU-process}, i.e. $y_\alpha \sim \mathcal{N}(\mu_\alpha, \sigma_\alpha^2/2 \nu_\alpha)$.
By comparison with Eq.~\eqref{eq: closure -- multivariate-OU}, the conditional mean is naturally defined as 
\begin{equation}
\label{eq: closure -- conditional-mean}
    \mu_\alpha(\bm x) =  \sum_{j, k = 1}^r \nu^{-1}_\alpha \left[ L^{21}_{\alpha j} x_j + Q^{211}_{\alpha j k} x_j x_k \right].
\end{equation}
Here we have omitted the contribution of the constant forcing term $\mathsfbi{F}^2$ so that the approximate fast process $\bm y$ preserves the zero-mean property of the POD coefficients for $\bm x = \bm 0$.
In this work we will make the simplifying assumption that the effective damping coefficients $\nu_\alpha$ are constant, although more generally they could be functions of $\bm x$.
Appropriate values for $\nu_\alpha$ can be determined from energy balance, as described below.

With this approximation, the average in Eq.~\eqref{eq: closure -- closed-ode} is a straightforward calculation resulting in the generalized Stuart-Landau model
\begin{subequations}
\begin{align}
    \hat{f}_i(\bm x) &= F^1_i + L^{11}_{ij} x_j + L^{12}_{i\alpha} \mu_\alpha(\bm x) + Q^{111}_{ijk} x_j x_k + 2 Q^{112}_{ij\alpha} x_j \mu_\alpha(\bm x) + \nu_\alpha^{-1} Q^{122}_{i \alpha \alpha} \sigma_\alpha \sigma_\alpha / 2 \\
    \label{eq: closure -- stuart-landau}
    &\equiv \hat{F}_i + \hat{L}_{ij} x_j + \hat{Q}_{ijk} x_j x_k + \hat{C}_{ijk\ell} x_j x_k x_\ell,
\end{align}
\end{subequations}
with the following closed quantities denoted by a hat:
\begin{subequations}
\label{eq: closure -- closed-tensors}
\begin{align}
    \hat{F}_i &= F^1_i + \nu_\alpha^{-1}  Q^{122}_{i\alpha \alpha} \sigma_\alpha \sigma_\alpha / 2 \\
    \hat{L}_{ij} &= L^{11}_{ij} + \nu^{-1}_\alpha
        L^{12}_{i\alpha} L^{21}_{\alpha j} \\
    \hat{Q}_{ijk} &= Q^{111}_{ijk} + \nu^{-1}_\alpha \left(
        L^{12}_{i\alpha} Q^{211}_{\alpha j k} + 2 Q^{112}_{ij\alpha} L^{21}_{\alpha k}
    \right) \\
    \label{eq: closure -- closed-tensors-cubic}
    \hat{C}_{ijk\ell} &= 2 \nu^{-1}_\alpha Q^{112}_{i j \alpha} Q^{211}_{\alpha k \ell}.
\end{align}
\end{subequations}
We will refer to the secondary reduction of a standard rank-$r$ POD-Galerkin model to the cubic model~\eqref{eq: closure -- stuart-landau} as the multiscale model reduction (MMR) approach to closure modeling.

\paragraph*{Interpretation of the cubic closure model}
The terms in the closure model are computed by summing the Galerkin tensors over the fast variables.
As a mnemonic for these averaged quantities, the superscripts for the partitioned system could be thought of as tensor contractions with the modification of the damping $\nu_\alpha^{-1}$.
For example, the cubic term $ Q^{112}_{i j \alpha} \nu^{-1}_\alpha Q^{211}_{\alpha k \ell}$ is suggestive of the slow variables ``filtering through'' the fast dynamics via the quadratic interactions $\mathsfbi{Q}^{211}$ and $\mathsfbi{Q}^{112}$.

Although the constant, linear, and quadratic terms are all modified as a result of the stochastic averaging procedure, the appearance of this cubic term is perhaps the most noteworthy.
It represents the leading-order contribution of the fast-slow nonlinear interaction in the slow dynamics due to the slow-slow interaction in the fast dynamics.
We will explore this in more detail in the following section, but it is a generalization of the weakly nonlinear Stuart-Landau mechanism~\citep{Stuart1958jfm, Landau1959book} that is capable of resolving the stabilizing influences of both mean flow deformation and the energy cascade.

The final ingredient in the multiscale closure model~\eqref{eq: closure -- stuart-landau} is the determination of the damping and diffusion coefficients $\nu_\alpha$ and $\sigma_\alpha$.
A reasonable approximation for the diffusion could be the mean of the neglected term
$
    \sigma_\alpha = \sum_{\beta} Q^{222}_{\alpha \beta \beta} \overline{y_\beta^2},
$
although in our numerical examples we find little difference with neglecting the diffusion altogether.
For the damping term we apply an energy balance condition on the closed model, derived from the assumption that the system is statistically stationary so that the average variation of kinetic energy vanishes~\citep{Noack2011book}:
\begin{equation}
\label{eq: closure -- energy-balance}
  \dv{t} \overline{x_i^2} = \overline{\hat{f}_i(\bm x) x_i } = 0.
\end{equation}
When the closed tensors~\eqref{eq: closure -- closed-tensors} are substituted for $\hat{\bm{f}}(\bm x)$, this simplifies to a linear system of equations for $\bm \nu^{-1}$.
The energy balance approximation does not require statistics of the fast variables, but due to the cubic term it does require well-converged fourth moments of the slow variables.
In particular, the vector $\bm \nu_{\rm inv}$ which is the element-wise inverse of $\bm \nu$ is the solution to the linear system
\begin{subequations}
\begin{equation}
    \mathsfbi{A} \bm{\nu}_{\rm inv} + \bm b = 0,
\end{equation}
\begin{align}
\begin{split}
  A_{i\alpha}     & = \sum_{j=1}^{r_0} L^{12}_{i\alpha} L^{21}_{\alpha j} \overline{a_i a_j}
    + \sum_{j,k=1}^{r_0} L^{12}_{i\alpha} Q^{211}_{\alpha j k} \overline{a_i a_j a_k} \\
              &\hspace{2cm}
    + 2 \sum_{j,k=1}^{r_0} Q^{112}_{i j \alpha} L^{21}_{\alpha k}  \overline{a_i a_j a_k}
    + 2 \sum_{j, k, \ell = 1}^{r_0} Q^{112}_{ij\alpha} Q^{211}_{\alpha k \ell} \overline{a_i a_j a_k a_\ell}
\end{split}
\\[2ex]
b_i &= L^{11}_{ii} \overline{a_i^2} + \sum_{j,k=1}^{r_0} Q^{111}_{ijk} \overline{a_i a_j a_k}
\end{align}
\end{subequations}

In general the damping should be positive since the linear stochastic approximation~\eqref{eq: closure -- multivariate-OU} and its diagonal simplification is only plausible if the mean $\bm \mu(\bm x)$ is linearly stable for all $\bm x$.
This is physically consistent with the energy cascade picture, in which we expect that energy will primarily flow ``downhill'' from the slow variables, representing large coherent structures and global instabilites, to the fast variables, representing the smaller, dissipative scales of the flow.
On this basis, negative values of the damping coefficients can also be set to zero to avoid introducing unphysical instabilities in the cubic term, for example.

We conclude the discussion of the application to Galerkin-type systems with a note on the computational scaling of the closure model.
While the simulation of the original linear-quadratic model is dominated by the quadratic term, which requires $\mathcal{O}(r^3)$ operations to evaluate, the cubic term in the closure model~\eqref{eq: closure -- stuart-landau} requires $\mathcal{O}(r_0^4)$ operations to evaluate.
In some cases this may mean that the cubic model is actually more expensive to simulate than the original Galerkin system, although whether or not this is the case will depend on the specific values of $r$, the dimension of the linear-quadratic system, and $r_0$, the dimension of the closed model.
For the examples given in Sec.~\ref{sec: results} either $r_0^4 < r^3$, as in Secs.~\ref{sec: results -- toy-problems},~\ref{sec: results -- cylinder} and~\ref{sec: results -- cav}, or the two are the same order of magnitude, as in Sec.~\ref{sec: results -- mixing-layer}.
As we will show, the primary advantage of this approach is not necessarily that it is faster to simulate than the usual POD-Galerkin system, but that it is more stable and physically faithful with many fewer modes.

\subsection{Relationship to invariant manifold modeling}
\label{sec: closure -- invariant}

Invariant manifold modeling is a method in which the fast variables are assumed to depend algebraically on the slow variables.
It is often applied near the threshold of bifurcation, when one may assume that there are only a small number of weakly unstable directions in the modal subspace and that the modal coefficients are typically small.

Under these assumptions, the fast variables can be approximated with a power series expansion with coefficients determined by a dynamical consistency condition~\citep{GuckenheimerHolmes, PavliotisStuart2012}.
For example, if the dynamics are close to a supercritical Hopf bifurcation then the leading-order approximation is a parabolic manifold that depends quadratically on the slow variables; replacing the fast variables with this approximation results in a cubic evolution equation for the slow variables, similar to Eq.~\eqref{eq: closure -- stuart-landau}.
This approach is closely related to the technique of center manifold reduction, which is typically applied to a system at the bifurcation point, yielding normal form-type amplitude equations for the evolution of the instability modes~\citep{Coullet1983, Carini2015jfm}.
Here we present the parabolic invariant manifold approximation to give a comparison with the stochastic averaging method, but we will conclude by showing that a different manifold model gives an equivalent result to Eq.~\eqref{eq: closure -- closed-tensors} for $\bm \sigma = \bm 0$.

We first shift the origin of the Galerkin system~\eqref{eq: galerkin -- linear-quadratic} to a fixed point~$\bm a^0$ defined by $\bm f(\bm a^0) = \bm 0$.
Defining fluctuations $\bm a'(t) = \bm a(t) - \bm a^0$, the Galerkin system is
\begin{subequations}
\begin{align}
\label{eq: closure -- galerkin-perturbation}
    \dot{a}_i' &= L_{ij}' a_j' + Q_{ijk} a_j' a_k'  \\
    L_{ij}' &= L_{ij} + (Q_{ijk} + Q_{ikj}) a^0_k.
\end{align}
\end{subequations}
We furthermore assume that the linear term $\mathsfbi{L}'$ is diagonalizable with an invertible set of eigenvectors $\mathsfbi{V}$ and eigenvalues $\bm \lambda = \mathrm{diag} (\bm{\Lambda})$, such that $\bm a' = \mathsfbi{V} \bm b$.
The similarity transform defined by $\mathsfbi{V}$ diagonalizes $\mathsfbi{L}'$ so that Eq.~\eqref{eq: closure -- galerkin-perturbation} becomes
\begin{subequations}
\begin{align}
    \dot{\bm{b}} &= \mathsfbi{V}^{-1} \bm{f}'(\mathsfbi{V} \bm b) \equiv \bm{f}^\Lambda(\bm b) \\
     &= \bm{\Lambda} \bm b + \mathsfbi{Q}^\Lambda (\bm b, \bm b), 
\end{align}
\end{subequations}
where $Q^\Lambda_{ijk} = V^{-1}_{i\ell} Q_{\ell m n} V_{mj} V_{nk}$ and the superscripts $\Lambda$ indicates that this tensor corresponds to the diagonalized coordinates $\bm b$.
As with the multiscale averaging method described in Sec.~\ref{sec: closure -- averaging}, we then partition the state $\bm b$ into slow variables $\bm x$ and fast variables $\bm y$, so that $\bm b = \begin{bmatrix} \bm x^T & \bm y^T \end{bmatrix}^T$ and $\bm{f}^\Lambda = \begin{bmatrix} \left(\bm f^{\bm x}\right)^T & \left(\bm f^{\bm y}\right)^T \end{bmatrix}^T$.
Similarly, the eigenvalues $\bm \lambda$ and quadratic tensor $\mathsfbi{Q}^\Lambda$ can be partitioned into $\mathsfbi{Q}^{111}$, $\mathsfbi{Q}^{112}$, etc., as before.

The central assumption of the invariant manifold model is that the fast variables can be expressed as an instantaneous function $\bm y \approx \bm h(\bm x)$ of the slow variables.
Dynamical consistency for the fast variables then requires that, to leading order,
\begin{subequations}
\begin{align}
    \dv{t} \bm y = \bm f^{\bm y}(\bm x, \bm h(\bm x)) &= \dv{t} \bm h(\bm x)\\
    \label{eq: closure -- consistency-jacobian}
    &= \mathbfcal{D}\bm h (\bm x) \bm f^{\bm x}(\bm x, \bm h(\bm x))
\end{align}
\end{subequations}
where $\mathbfcal{D}\bm h (\bm x)$ is the Jacobian of $\bm h$ evaluated at $\bm x$~\citep{GuckenheimerHolmes}.

Assuming all of the state variables in $\bm b$ are typically small, the manifold equation can be expanded as a power series in $\bm x$.
Preservation of the fixed point at the origin requires $\bm h(\bm 0) = \bm 0$, and in the case of a center manifold reduction where the real parts of the slow eigenvalues vanish, the condition that the $\bm h(\bm x)$ be tangent to the center manifold at the origin also implies that $\mathbfcal{D}\bm h(\bm 0) = \bm 0$.
Under these conditions the first nontrivial terms in the expansion are quadratic and the expansion coefficients can be determined by matching powers of $\bm x$ in the consistency condition~\eqref{eq: closure -- consistency-jacobian}.

The connection between this method and stochastic averaging can be made clear by viewing $\bm y(t) = \bm h(\bm x, t)$ as an explicit function of time and rewriting the dynamical consistency condition as a Koopman generator acting on the $\alpha^\mathrm{th}$ component of the manifold equation:
\begin{equation}
\label{eq: closure -- consistency-advection}
    \pdv{h_\alpha}{t} = \bm{f}^{\bm x}(\bm x, \bm h(\bm x)) \cdot \nabla h_\alpha.
\end{equation}
Since the observable in such an equation is arbitrary, this is equivalent to the averaged backwards Kolmogorov equation~\eqref{eq: closure -- closed-bke-rescaled} if
\begin{equation}
    f_i^{\bm x}(\bm x, \bm h(\bm x)) = \int_{\mathcal{Y}} \rho_{\bm x}^\infty(\bm y) f_i^x(\bm x, \bm y) \dd \bm y.
\end{equation}
In the limit of zero diffusion in the fast process, the stationary distribution approaches a Dirac delta function $\rho_{\bm{x}}^\infty = \delta^{r-r_0}(\bm y - \bm \mu(\bm x))$ and the two are equivalent with $\bm h(\bm x) = \bm \mu(\bm x)$.
With the diagonal drift approximation~\eqref{eq: closure -- diagonal-drift} and constant damping $\bm \nu$, the zero-diffusion limit of the proposed Galerkin closure in Sec.~\ref{sec: closure -- galerkin} can be seen as an invariant manifold model with $\bm h(\bm x)$ defined as the conditional mean~\eqref{eq: closure -- conditional-mean}.
Because we derived the stochastic averaging model with the assumption of real-valued coefficients and the parabolic model uses the complex-valued eigenvector coordinates, it is less obvious that the parabolic invariant manifold corresponds to a particular choice of $\bm \nu$ and $\bm \mu$ in Eq.~\eqref{eq: closure -- diagonal-drift}.


Although the invariant manifold reduction is more theoretically straightforward than the stochastic averaging closure, it generally requires stronger assumptions on the system.
In particular, the development of the parabolic manifold model requires that the origin be a fixed point and that the real part of the slow eigenvalues is zero.
This does not raise any difficulties for simple weakly unstable configurations where the Galerkin model preserves fixed points of the fluid flow, or if the steady state is used as the base of the modal expansion.
However, when the flow is far from a bifurcation point, such as for a fully turbulent flow, the mean flow is often a more relevant base of the expansion. 
In this case the resulting Galerkin models may not necessarily have fixed points that coincide with meaningful states of the underlying system.
In addition, the invariant manifold reduction relies on smallness of the coefficients (justifying the power series expansion) and the assumption that the higher-order modes can be expressed as instantaneous algebraic functions of the active coefficients.
For chaotic or turbulent dynamics neither of these is necessarily justified.

\subsection{Finite-time thermodynamics}
\label{sec: closure -- ftt}

Before concluding the discussion of the multiscale closure model, we give a brief comparison to the finite-time thermodynamics (FTT) framework introduced by~\citet{Noack2008jnet}.
This approach begins from the linear-quadratic POD-Galerkin system, Eq.~\eqref{eq: galerkin -- linear-quadratic}, but rather than assuming that the unresolved variables are the higher-order modal coefficients, the FTT model decomposes the modal coefficients $\bm a(t) \in \R^r$ into an ``ensemble-averaged'' value $\bm x(t) \in \R^r$ and a ``fluctuating'' value $\bm y(t) \in \R^r$ in the spirit of the Reynolds decomposition.
As a result, both the resolved and unresolved variables correspond to the \emph{same} spatial modes, and the modal expansion~\eqref{eq: modal-ansatz} becomes
\begin{equation}
    \bm u(\bm x, t) \approx \bm \psi_0(\bm x) + \sum_{i=1}^r ( x_i(t) + y_i(t) ) \bm \psi_i(\bm x).
\end{equation}

The second major difference is the assumption of the behavior of the unresolved variables $\bm y$ on ensemble averaging.
The ensemble average is with respect to the probability distribution $\rho$ of the fluctuations $\bm y$.
The MMR method assumes that this distribution is parameterized explicitly by the slow variables $\bm x$, so that $\rho = \rho(\bm y; \bm x(t))$ (written as $\rho_{\bm x}^\infty(\bm y)$ in Sec.~\ref{sec: closure -- averaging}).
On the other hand, the Reynolds-averaging analogy in FTT leads to the assumption that the distribution $\rho$ may be time-varying but does not depend on $\bm x$, so that $\rho = \rho(\bm y; t)$.

\begin{table}
\centering
  \begin{tabular}{ || c  c  c ||}
 \hline
    & MMR & FTT  \\
 \hline
 \hline
   $\rho$ & $\rho(\bm y; \bm x)$ & $\rho(\bm y; t)$ \\
 \hline
   $\E[y_i]$ & $\mu_i(\bm x)$ & 0 \\
 \hline
   $\E[y_i^2]$ & $\sigma_i^2$ & $2E_i(t)$ \\
 \hline
  \end{tabular}
  \caption{
  \small
  \textbf{Statistics of unresolved variables for the multiscale and finite-time thermodynamics models.}
  MMR assumes the distribution is parameterized by the slow variables $\bm x$, while FTT assumes it is independent and zero-mean, although with possibly fluctuating variance.
  We have found the variance $\sigma_i^2$ to be negligible in MMR if it is assumed constant, but more generally it could be also modeled as a function of $\bm x$.
  Note that the unresolved variables $\bm y$ are also defined differently between the two methods (see Sec.~\ref{sec: closure -- ftt}).
  }
  \label{tab: closure -- mmr-vs-ftt}
\end{table}

By definition, the fluctuating amplitudes $\bm y$ have zero mean with respect to this distribution, so that
\begin{equation}
\label{eq: closure -- ftt-average}
    \E[y_i] \equiv \int_\mathcal{Y} y_i \rho(\bm y; t) \dd \bm y = 0, \qquad i = 1, 2, \dots r.
\end{equation}
Here we use the notation $\E[\cdot]$ to denote the ensemble average with respect to the unknown distribution $\rho(\bm y, t)$ to distinguish it from the usual time average of observed values used in the rest of this work.

As a result of Eq.~\eqref{eq: closure -- ftt-average}, the mean and fluctuating amplitudes are linearly uncorrelated with respect to this average:
\begin{equation}
    \E[x_i y_j] = 0, \qquad i, j = 1, 2, \dots r.
\end{equation}
The normalization of the probability distribution also implies that $\E[x_i] = x_i$.
Finally, FTT assumes that the fluctuations themselves are uncorrelated, with the usual variance-kinetic energy relationship for POD coefficients.
Combining these assumptions, the second-order statistics of the POD coefficients with respect to the ensemble average defined by $\rho(\bm y; t)$ are
\begin{equation}
\label{eq: closure -- ftt-variance}
    \E[a_i a_j] = x_i x_j + 2 E_i(t) \delta_{ij},
\end{equation}
where $E_i(t) = \E[y_i^2]/2$.
A comparison of the assumptions on the distributions is shown in Table~\ref{tab: closure -- mmr-vs-ftt}.
Note that since the variance $E_i$ must be non-negative, Eq.~\eqref{eq: closure -- ftt-variance} implies that the statistics of $\bm a$ and $\bm x$ are not identical unless $\bm E(t) = 0$ for all time.

Expanding the POD coefficients into mean and fluctuating component and taking the ensemble average gives a set of $2r$ evolution equations for the mean values $\bm x$ and the fluctuation kinetic energy $\bm E$:
\begin{subequations}
\label{eq: closure -- ftt}
\begin{align}
    \dot{x}_i &= \sum_{j=1}^r L_{ij} x_j + \sum_{j, k=1}^r Q_{ijk} x_j x_k + 2 \sum_{j=1}^r Q_{ijj} E_j \\
    \dot{E}_i &=  2 \sum_{j=1}^r\left( L_{ii} + 2 Q_{iij} x_j \right) E_j + \sum_{j, k=1}^r Q_{ijk} \E[y_i y_j y_k].
\end{align}
\end{subequations}
The remainder of the FTT method centers on deriving a nonlinear closure model for the third-order statistics $\E[y_i y_j y_k]$; see~\citet{Noack2008jnet} for details.

The main similarity between the FTT and multiscale methods is that both seek to approximate the effect of unresolved fluctuations on the resolved variables, resulting in nonlinear closure models.
However, the two are philosophically different in that MMR assumes the unresolved variables correspond to different spatial structures and evolve on faster time scales, while FTT models them as linearly uncorrelated fluctuations associated with the same spatial modes as the resolved variables.

\citet{Noack2008jnet} showed that the FTT model is quite successful at predicting post-transient energy levels in several flow configurations, including the cylinder wake and a turbulent homogeneous shear flow.
This method also has the advantage that it does not assume any separation of scales in amplitude or time.
On the other hand, there are some theoretical difficulties in working with the FTT framework, including that Eq.~\eqref{eq: closure -- ftt-variance} muddies the relationship between the second-order statistics of the modal coefficients $\bm a$ and $\bm x$, which are foundational to the POD method.
FTT also increases the state dimension by a factor of two and cannot be used to derive the generalized Stuart-Landau equations that directly account for effects such as mean flow deformation and the energy cascade.
However,~\citet{Schlegel2009},~\citet{Noack2011book}, and~\citet{Cordier2013} showed that a simplified nonlinear closure model based on Eq.~\eqref{eq: closure -- ftt} has been shown to stabilize POD-Galerkin models of a mixing layer similar to that examined in Sec.~\ref{sec: results -- mixing-layer} without increasing the state dimension (see App.~\ref{app: closure}).

%% file: S5_Numerics.tex
\section{Flow configurations}
\label{sec: numerical}

In this work we consider models of three fluid flows: the canonical flow past a cylinder at Reynolds number 100, a lid-driven cavity flow, and an incompressible mixing layer with two domain extents.
We construct projection-based models of these flows based on direct numerical simulation (DNS) using the open-source Nek5000 spectral element solver~\citep{Nek5000}.
The semi-implicit time-stepping integrates diffusive terms with third order backwards differentiation and convective terms with a third order extrapolation.

Once the DNS is complete, we estimate the POD modes and coefficients using the \texttt{modred} library, which provides a set of parallelized, high-performance algorithms in Python for linear modal decompositions and model reduction~\citep{Belson2014modred}.
We then compute the components of the Galerkin system~\eqref{eq: galerkin -- linear-quadratic} by extracting the weight matrix and necessary gradients from the POD modes with the built-in Nek5000 post-processor.
Both the linear-quadratic Galerkin model~\eqref{eq: galerkin -- linear-quadratic} and the cubic closure~\eqref{eq: closure -- stuart-landau} are simulated using the LSODA time-stepping algorithm available from \texttt{scipy} with analytic Jacobian matrices for each system.

\subsection{Flow past a circular cylinder}
\label{sec: numerical -- cyl}

\begin{figure}
\centering
\includegraphics[width=0.935\textwidth]{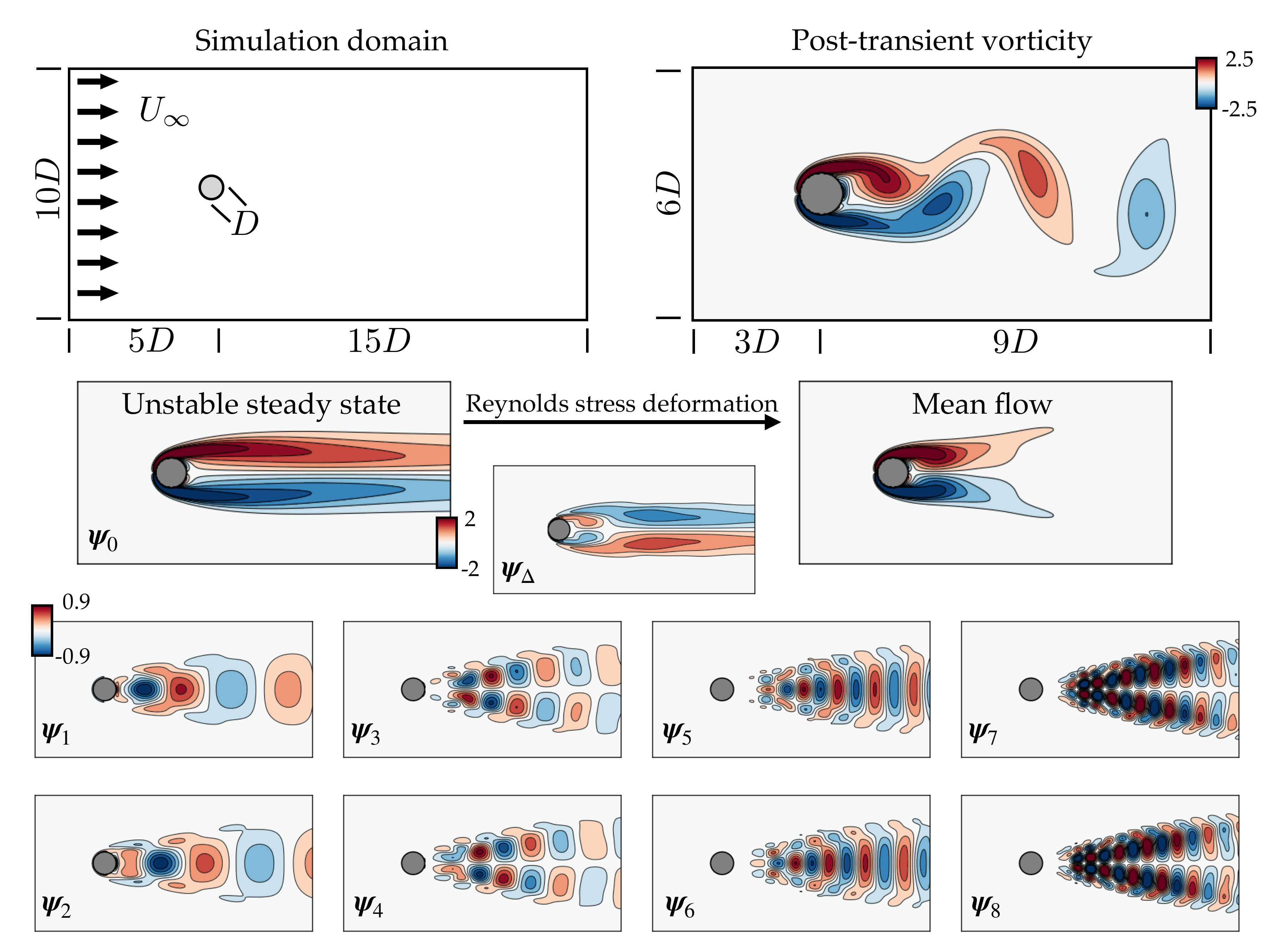}
\caption{
{\small
\textbf{Vortex shedding in the wake of a circular cylinder at $\Rey = 100$.}
En route to the periodic post-transient flow, Reynolds stresses deform the background flow from the unstable steady state to the marginally stable mean flow.
This can be approximated in a model by augmenting the POD basis with the ``shift mode'' $\bm \psi_\Delta$.
The usual POD modes show the typical structure of spatial and temporal harmonics describing periodically advecting flow features.
}}
\label{fig: numerical -- cyl-config}
\end{figure}

The vortex shedding in the wake behind a cylinder at Reynolds number 100, based on the free-stream velocity and cylinder diameter, is a canonical flow configuration for reduced-order modeling~\citep{Noack2003jfm} and it is shown in Fig.~\ref{fig: numerical -- cyl-config}.
We simulate this flow on a domain of 2600 sixth-order spectral elements on $x, y \in (-5, 15) \times (-5, 5)$ refined close to the cylinder wall; further details and analysis can be found in~\citet{Loiseau2018,Loiseau2018jfm}.
We first perform a global stability analysis of the the unstable steady state, determined by solving the stationary Navier-Stokes equations with selective frequency damping~\citep{Akervik2006pof}, using a Krylov-Schur time-stepping algorithm~\citep{Loiseau2019book}.
The transient simulation is then initialized with the unstable steady state perturbed by the least-stable global eigenmode, normalized so that its energy is $10^{-4}$.

The flow regime at Reynolds number 100 is beyond the Hopf bifurcation leading to vortex shedding but below the onset of three-dimensional instability~\citep{Noack1994pof, Barkley1996jfm, Ma2002jfm}.
The flow is also far enough from the threshold of bifurcation that weakly nonlinear analyses fail to predict the amplitude and frequency of the vortex shedding limit cycle, although a variety of techniques have been used to derive accurate models for the transient and post-transient wake~\citep{Noack2003jfm, Noack2011book, ManticLugo2014prl, Loiseau2017jfm, Loiseau2018jfm}.

Following~\citet{Noack2003jfm} and~\citet{Loiseau2018}, we compute POD modes from 100 snapshots of the mean-subtracted velocity field, equally spaced over one period of vortex shedding; the pressure gradient term can shown to be negligible for Galerkin models.
In order to construct a basis that can span the deformation between the unstable steady state and mean flow, we augment the POD basis with the ``shift mode'', visualized in Fig.~\ref{fig: numerical -- cyl-config}.
This additional mode is computed from the difference between the base and mean flows and then orthonormalized with respect to the rest of the POD modes with the Gram-Schmidt process.
We denote the coefficient of this mode by $a_\Delta(t)$ to distinguish it from the usual POD modes.

The full expansion basis then consists of eight POD modes capturing the limit-cycle vortex shedding, plus the shift mode.
We take the unstable steady state to be the base flow of the modal expansion so that the origin is a fixed point of the POD-Galerkin system.
This basis accurately represents the kinematics of the post-transient flow, with a normalized energy residual of $\mathcal{O}(10^{-4})$~\citep{Loiseau2018}.

\subsection{Lid-driven cavity flow}
\label{sec: numerical -- cav}

\begin{figure}
\centering
\includegraphics[width=0.935\textwidth]{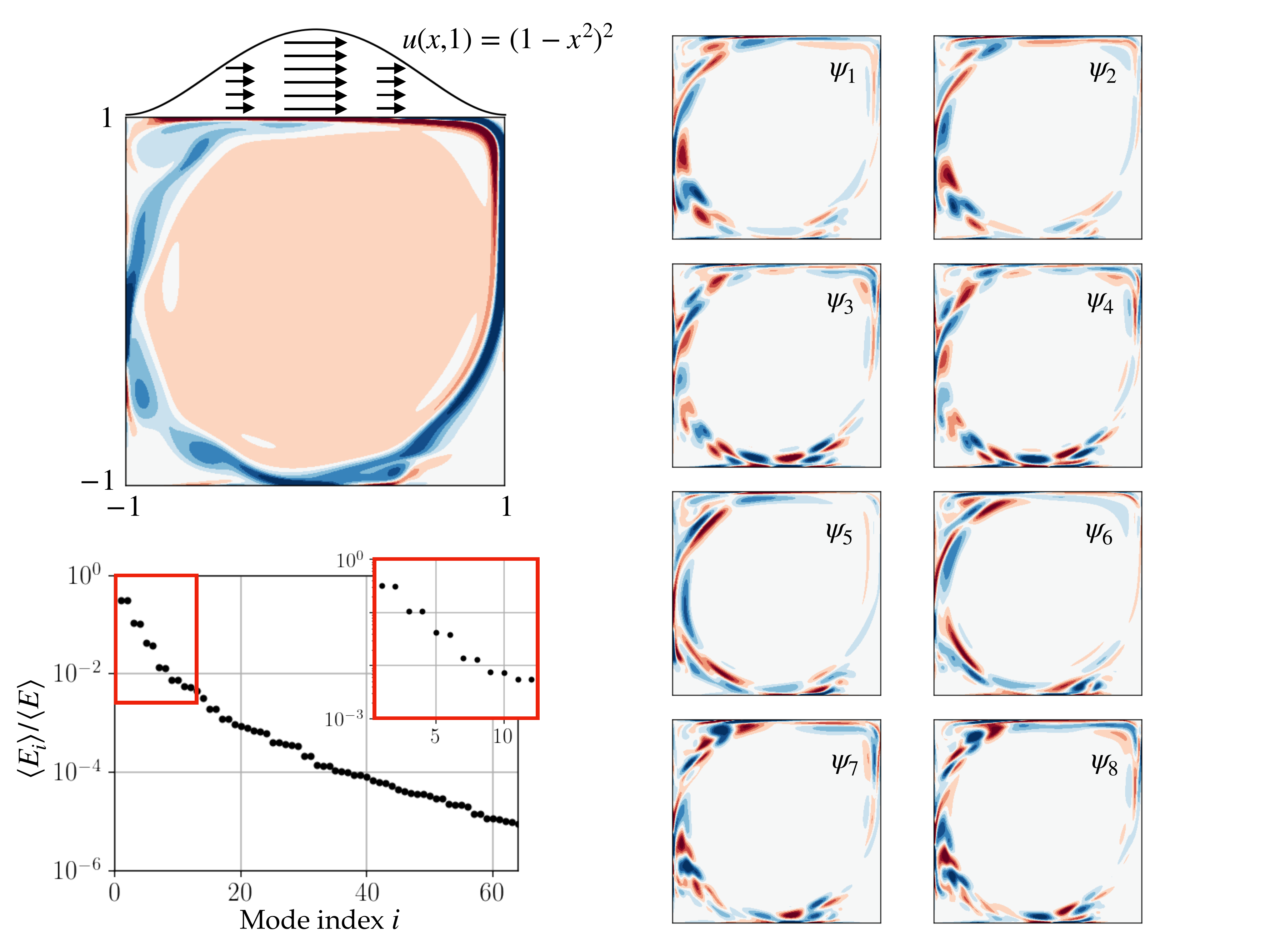}
\caption{
{\small
\textbf{Regularized lid-driven cavity flow at $\Rey = 20,000$.}
The singular value spectrum is slow to converge (left), indicating that relatively many modes are necessary for an accurate kinematic approximation.
The leading POD modes themselves appear roughly in pairs with similar spatial structure and frequency content.
As for the cylinder wake shown in Fig.~\ref{fig: numerical -- cyl-config}, this usually indicates oscillatory dynamics, although the evolution of the temporal coefficients is much more irregular in this chaotic flow than for the periodic wake.
} }
\label{fig: numerical -- cav-config}
\end{figure}

Lid-driven cavity flow is another idealized geometry that serves as a benchmark problem for numerical methods and model reduction~\citep{Arbabi2017prf, Arbabi2019, Lee2019cavity, Rubini2020jfm}.
The flow, shown in Fig.~\ref{fig: numerical -- cav-config}, takes place on a square domain with $(x, y) \in (-L, L) \times (-L, L)$ with $L=1$.  A velocity profile is imposed on the upper boundary, given by
\begin{equation}
    u(x, L) = U(1 - x^2)^2.
\end{equation}
This driving velocity profile roughly approximates the shear-driven flow over an open cavity at a much lower computational cost, although the dynamics of the shear- and lid-driven cavities are different.
The regularized fourth-order velocity profile on the lid avoids numerical problems where the lid meets the no-slip boundaries on the other walls.

The standard Reynolds number for this flow is defined as $\Rey = UL/\nu$ based on the cavity half-height and forcing velocity.
The flow follows the stereotypical Ruelle-Takens route to chaos parameterized by Reynolds number, undergoing a Hopf bifurcation at $\Rey_c^{(2)} \approx 10,250$, a secondary Hopf bifurcation to quasiperiodic flow at $\Rey_c^{(2)} \approx 15,500$, and a final bifurcation to chaos near $\Rey_c^{(3)} \approx 18,000$~\citep{Lee2019cavity}.
We simulate the flow in the chaotic regime at $\Rey = 20,000$.

The numerical configuration is similar to that described by~\citet{Arbabi2019}.
The domain is discretized with 50 seventh-order spectral elements in each direction, refined towards the walls.
The flow is integrated for 3000 time units, saving snapshots at a sampling rate $\Delta t = 0.1$.
Again, the pressure gradient term vanishes for this closed flow~\citep{Noack2005jfm}, so the velocity POD modes are computed from the first 5000 snapshots, with the remaining 25000 snapshots retained for the statistical comparison in Sec.~\ref{sec: results -- cav}.
Figure~\ref{fig: numerical -- cav-config} shows a snapshot of the DNS along with the leading POD modes and the singular value spectrum.

\subsection{Incompressible mixing layer}
\label{sec: numerical -- mix}

\begin{figure}
\centering
\begin{overpic}[width=\linewidth]{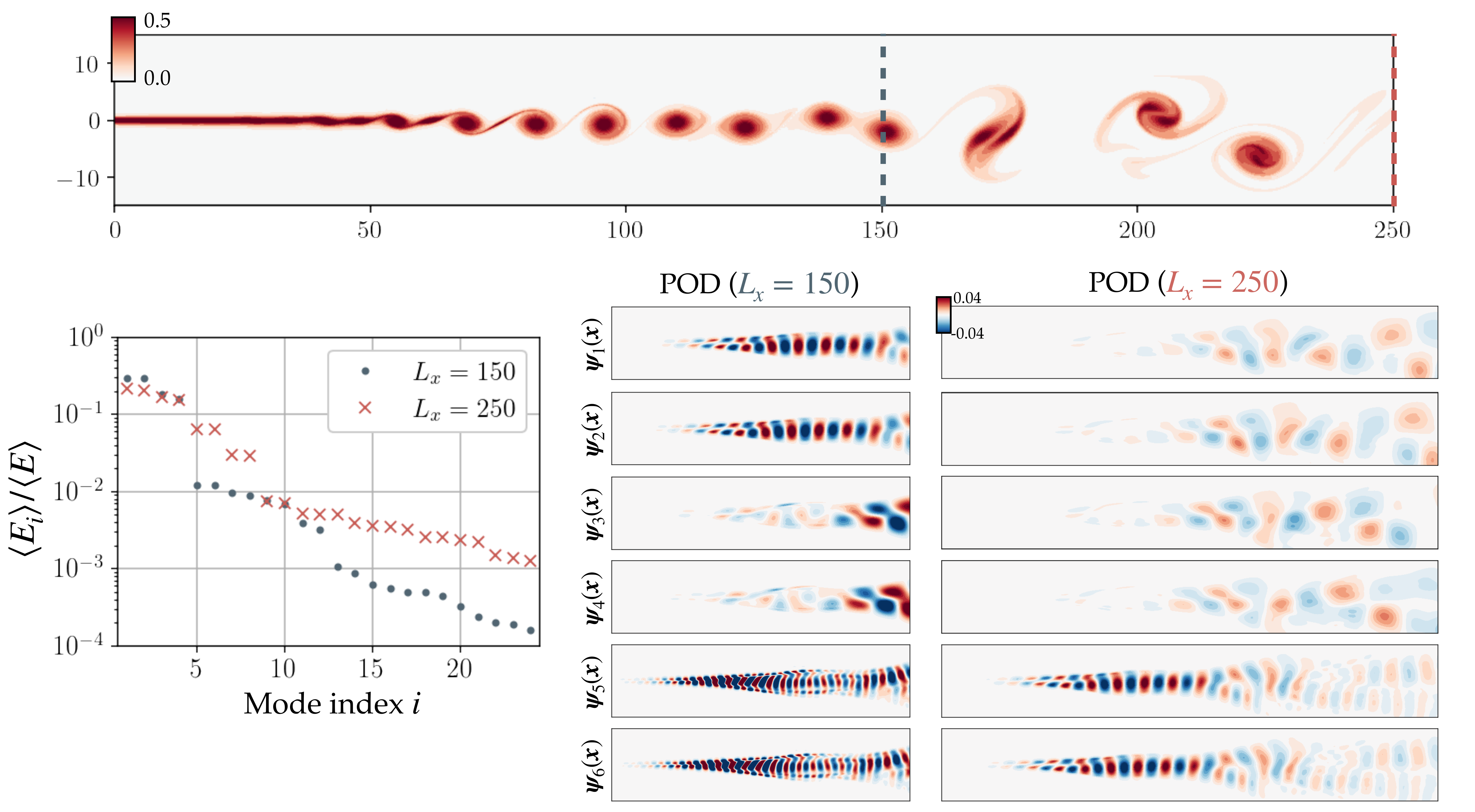}
\end{overpic}
\caption{
{\small
\textbf{Proper orthogonal decomposition (POD) applied to the mixing layer.}
The primary shear layer instability forms Kelvin-Helmholtz waves that roll up into vortices, which in turn undergo successive vortex pairing (top).
Modes computed on both the short and long domains reveal modes related to the dominant flow features: the shear layer instability and the downstream vortex pairing.
Although the vortex pairing is secondary to the upstream instability, on the longer domain it accounts for a larger proportion of the fluctuation kinetic energy.
Higher-order modes not pictured here include harmonics, nonlinear crosstalk, and modes related to irregularity in the location of the vortex pairing events.
}}
\label{fig: numerical -- mix-modes}
\end{figure}

As a more difficult test of the proposed MMR closure method, we also examine models of an incompressible mixing layer, pictured in Fig.~\ref{fig: numerical -- mix-config}.
The mixing layer is a canonical example of a free shear flow exhibing convective instability~\citep{Monkewitz1982, Huerre1985, Huerre1990arfm}.
Inlet forcing is used to excite Kelvin-Helmholtz instability waves, which roll up into approximately discrete vortices.
These vortices are subject to a secondary instability~\citep{Brancher1997prl} and undergo successive helical pairing at the subharmonic of the primary Kelvin-Helmholtz frequency, a process that drives the linear growth of the mixing layer~\citep{Winant1974}.
At higher Reynolds numbers and in three dimensions, the turbulent mixing layer is dominated by the linear growth of coherent structures similar to those seen in 2D simulation~\citep{Brown1974, Stanley1997}, which play a major role in mixing, transport, and entrainment of the turbulent flow~\citep{Hussain1981}.
From a global perspective, the flow behaves as an amplifier, where strong non-normality in the linear operator results in transient algebraic energy growth of small perturbations~\citep{Chomaz2005arfm}.

By analogy to numerical methods, modeling an advection-dominated flow with a POD-Galerkin approach is generally ill-advised in much the same way that pseudospectral methods are not ideal for solving hyperbolic problems.
Although the vortices shed in the flow past a cylinder are also traveling waves in a sense, the Kelvin-Helmholtz instability in the mixing layer is convective in nature as opposed to the absolute instability in the cylinder wake.
For these reasons, Galerkin projection onto a global modal basis such as POD is not a natural way to approximate shear flows like the mixing layer.
Nevertheless, both free and wall-bounded shear layers are present in a wide variety of flows, so in order to employ a low-dimensional dynamical systems model it is important that it be able to capture these features.
The challenge of modeling advection-driven phenomena in a global reduced-order modeling framework motivates the use of similar configurations in several studies of stabilization methods for Galerkin-type systems~\citep{Ukeiley2001, Noack2005jfm, Cordier2013, Balajewicz2013jfm}.

Since the flow becomes increasingly complex downstream, the difficulty of modeling the flow can be tuned by varying the streamwise extent of the domain. 
We construct models on both short ($L_x = 150$) and long ($L_x = 250$) domains, as shown in Fig.~\ref{fig: numerical -- mix-modes}.
Over the extent of the short domain, the flow is mainly characterized by the initial instability and vortex rollup, with the earliest signs of vortex pairing at $x \gtrsim 130$.
On the full domain, the vortex pairing accounts for a much larger fraction of the fluctuation kinetic energy, as shown by the POD analysis below.

The inlet profile is a standard hyperbolic tangent:
\begin{equation}
\label{eq: numerical -- tanh}
    U(y) = \bar{U} + (\Delta U / 2 ) \tanh 2y,
\end{equation}
where $\bar{U} = (U_1 + U_2)/2 \equiv 1$ and $\Delta U = U_1 - U_2$ in terms of the free-stream velocities $U_1$ and $U_2$ above and below the layer, respectively, and the length scale is nondimensionalized by the initial vorticity thickness
$
    \delta_\omega = {\Delta U}/{U'(0)}.
$
We define a Reynolds number $\Rey = \delta_\omega(0) \Delta U / \nu$ based on intial vorticity thickness, velocity difference $\Delta U$ across the layer, and kinematic viscosity $\nu$ and set $\Rey = 500$ in the simulation.
The domain consists of 5200 eleventh-order spectral elements on $x, y \in (-10, 300) \times (-50, 50)$, but the full streamwise extent is masked to $x, y \in (0, 250) \times (-20, 20)$ for modeling in order to discount any boundary effects.
The domain is shown in Fig.~\ref{fig: numerical -- mix-config}.

This configuration roughly follows that described by~\citet{Balajewicz2013jfm, Cordier2013}, with two key differences.
First, we simulate incompressible flow, while these authors simulate isothermal compressible flow at low Mach numbers ($\mathrm{Ma} \sim 0.1$).
Consequently, we replace the ``sponge'' boundary conditions for far-field acoustic waves with a region of increased viscous damping corresponding to $\Rey = 50$ for $x > 250$ to avoid numerical instability at the outflow.
Second, rather than disturbing the inlet with random solenoidal perturbations, we use eigenfunction forcing described in App.~\ref{app: stability}, similar to that employed by~\citet{Colonius1997}.

The DNS is run until a final time of $t=2820$, corresponding to fifty periods of the lowest-frequency component of the eigenfunction forcing, saving at every tenth time step with $\Delta t = 0.00705$.
The POD modes are then computed from the first 10\% of these snapshots (4000 fields), with the remainder reserved for statistical comparison with the models.
The domain truncation can be accomplished by simply setting elements of the weight matrix $\mathsfbi{W}$ in Eq.~\eqref{eq: galerkin -- weighted-inner-product} to zero for mesh locations with $x > L_x$  so that these locations do not contribute to the statistics in the correlation matrix.
For the mixing layer we compute modes with both velocity and pressure fields, as described in Sec.~\ref{sec: galerkin -- pod}, since the pressure gradient term is not necessarily negligible in Galerkin models of free shear flows~\citep{Noack2005jfm}.

Figure~\ref{fig: numerical -- mix-modes} shows the leading POD eigenvalues and modes for both domains.
In both cases the bulk of the fluctuation kinetic energy is contained in the first four modes, with the remainder of the spectrum decaying relatively slowly.
As expected, the dominant mode pair for the shorter domain corresponds to the Kelvin-Helmholtz waves, while the second pair represents the onset of vortex pairing at roughly half the spatial wavenumber.
Higher-order modes are either harmonics (such as the third mode pair) or less regular downstream structures related to slight variations in the vortex pairing.

The directed graph of energy transfers in Fig.~\ref{fig: intro -- overview} is constructed from the modes computed on the short domain.
Specifically, the leading mode pair and its first three harmonics are shown for a straightforward visualization of the energy cascade; the nodes labeled $a_1-a_8$ thus actually correspond to mode pairs $(\bm \varphi_1, \bm \varphi_2)$, $(\bm \varphi_5, \bm \varphi_6)$, $(\bm \varphi_{17}, \bm \varphi_{18})$, and $(\bm \varphi_{24}, \bm \varphi_{25})$.
The thickness of the edge from node $a_i$ to $a_j$ is proportional to $\sum_{k} Q_{ijk} \overline{a_i a_j a_k}$ for the quadratic Galerkin tensor $\mathsfbi{Q}$, which is roughly related to average nonlinear energy transfer from one mode to the other, although it should not necessarily be interpreted in any rigorous way beyond the purposes of conceptual visualization of the energy cascade.

On the longer domain the vortex pairing makes up a larger proportion of the fluctuation kinetic energy.
Consequently, the leading four modes are spatially localized downstream, representing vortex pairing, while the upstream linear instability is not evident until the third mode pair, even though the vortex pairing is secondary to the shear layer instability from a physical perspective.

The slowly decaying POD spectrum and wavelike spatial structure of the modes are a consequence of the advection-driven nature of this flow.
It is well known that the space-time separation of variables assumed by POD is not a natural representation of traveling waves.  
This is one reason that free shear flows have long posed difficulties for POD-Galerkin modeling, making this a challenging test case for the MMR approach.

%% file: S6_Results.tex
\section{Results}
\label{sec: results}

In this section we evaluate the proposed multiscale model reduction (MMR) closure method on several numerical test problems.
Before modeling the flows described in Sec.~\ref{sec: numerical}, we first explore the capability of MMR to resolve stereotypical subscale stabilization mechanisms in simple analytically tractable model problems in Sec.~\ref{sec: results -- toy-problems}, namely for the mean-field model of \citet{Noack2003jfm} and for Burgers' equation.

In Sec.~\ref{sec: results -- cylinder} we analyze vortex shedding behind a circular cylinder, showing that the method reproduces the classic invariant manifold model due to~\citet{Noack2003jfm}, which must account for both mean flow deformation and the energy cascade in order to accurately model the transient dynamics.
As a more challenging test in Sec.~\ref{sec: results -- cav}, we show that the multiscale closure stabilizes a reduced-order model of the chaotic lid-driven cavity flow, while closely matching the spectral energy content of the DNS.
Finally, in Sec.~\ref{sec: results -- mixing-layer} we develop a model for convective instability and subharmonic vortex pairing in the spatially evolving mixing layer.

By means of these examples, we show that the proposed approach for multiscale closure is a systematic method for eliminating degrees of freedom from reduced-order models that can resolve critical subscale phenomena and produce stable models of dynamically complex unsteady flows.
More broadly, the multiscale framework represents a general description of the means by which effective cubic nonlinearity arises in low-dimensional models.

\subsection{Analytically tractable model problems}
\label{sec: results -- toy-problems}

There are two principal mechanisms by which nonlinear interactions with higher-order modes effectively stabilize the flow: Reynolds stress-induced mean flow deformation and the energy cascade.
Before demonstrating the MMR closure introduced in Sec.~\ref{sec: closure} on DNS of unsteady fluid flow, we first confirm that it can resolve both of these effects by examining simple representative analytic models.

\subsubsection{Nonlinear stability in a mean-field model}
\label{sec: results -- mean-field}
Nonlinear stability analysis was born out of the observation that the exponential energy growth predicted by linear theory when the flow has at least one unstable eigenvalue is untenable at long times. 
Furthermore, the key assumption of infinitesimal perturbation amplitudes is violated once the instability grows to finite amplitude.
It is conceivable that the energy growth eventually leads to a breakdown into turbulence, but it was observed experimentally (e.g.~\citet{Taylor1923}) that this is not always the case.

Instead, if the perturbations are assumed to be of small but finite amplitude, the linear stability problem can be considered the first term in an asymptotic expansion.
The next leading-order terms suggest that the Reynolds stresses from the nonlinear self-interaction of the perturbation tends to deform the slowly-varying background flow from the unstable steady-state towards a neutrally stable mean flow~\citep{Stuart1958jfm}.
This type of weakly nonlinear analysis has since been applied to a wide range of flows and bifurcation scenarios~\citep{Sipp2007jfm, Meliga2009jfm, Meliga2011jfm, Rigas2017jfm}.
Although this has proven to be a highly successful approach, it is only strictly applicable near the threshold of bifurcation and requires invasive access to the various linear operator ``building blocks" of the governing equations.
The self-consistent mean flow modeling framework~\citep{ManticLugo2014prl, ManticLugo2016jfm, Meliga2017jfm, Rigas2020jfm} is more general in terms of parametric dependence, but it does not lead to a low-dimensional dynamical system.

Resolving mean flow deformation has been shown to be a key ingredient in stabilizing reduced-order models.
Models that only include POD modes derived from expansion around the mean give accurate approximations on short time horizons, but they eventually tend towards structural instability~\citep{Deane1991pof, Ma2002jfm}.
However, explicitly incorporating the stabilizing feedback mechanism via Reynolds stress modeling~\citep{Aubry1988jfm, Holmes1996}, an augmented POD basis~\citep{Noack2003jfm, Luchtenburg2009jfm, Noack2011book}, or a normal form ansatz~\citep{Deng2020jfm} in models of globally unstable flows, has significantly improved their robustness.
Here we show that if the POD basis spans the mean flow deformation, the proposed multiscale model reduction method can resolve its stabilizing effect without introducing additional degrees of freedom.
This is consistent with the invariant manifold reduction approach of~\citet{Noack2003jfm}, but does not require treating the mean flow deformation differently than any of the other unresolved fast variables.

A classic illustrative model of the nonlinear stability mechanism is the three-dimensional ``mean-field'' system proposed by~\citet{Noack2003jfm}.
The model consists of a pair of unstable slow variables $a_1$ and $a_2$ along with the fast variable $a_\Delta$:
\begin{subequations}
\label{eq: analytic -- mean-field}
\begin{align}
    \dot{a}_1 &= \lambda a_1 - \omega a_2 - a_1 a_\Delta \\
    \dot{a}_2 &= \omega a_1 - \lambda a_2 - a_2 a_\Delta \\
    \dot{a}_\Delta &= -a_\Delta + \mu (a_1^2 + a_2^2).
\end{align}
\end{subequations}
The slow variables represent amplitudes of a pair of global instability modes, while $a_\Delta$ models a background deformation that tends to reduce the linear growth rate.
This system is a simplified model with similar behavior to the cylinder wake discussed in Sec.~\ref{sec: results -- cylinder}.

In many cases the growth rate $\lambda$ is relatively small, so that the standard approach is to use the adiabatic elimination approximation $\dot{a}_\Delta = 0$.
The fast variable is then given by an algebraic relationship that pins the system to the parabolic invariant manifold $a_\Delta \approx \mu (a_1^2 + a_2^2) $.
Replacing $a_\Delta$ in Eq.~\eqref{eq: analytic -- mean-field} results in a cubic Stuart-Landau equation for the complex amplitude $A(t) = a_1 + i a_2$:
\begin{equation}
\label{eq: analytic -- stuart-landau}
    \dot{A} = (\lambda + i \omega) A - \mu A |A|^2.
\end{equation}

We recapitulate these results in order to show that this mechanism is also resolved by the multiscale closure~\eqref{eq: closure -- stuart-landau}.
Here the fast dynamics are already linear in $a_\Delta$ with damping constant $\nu = 1$.
The fast-slow linear blocks $\mathsfbi{L}^{12}$ and $\mathsfbi{L}^{21}$ are also zero, so the only modification due to the closure model is the cubic term $\hat{\mathsfbi{C}}$.
The only nonzero entries in the quadratic tensor are $Q_{1 1 \Delta} = Q_{2 2 \Delta} = -1$ and $Q_{\Delta 1 1} = Q_{\Delta 2 2} = \mu$, leading to cubic terms $\hat{C}_{1111} = \hat{C}_{1122} = \hat{C}_{2211} = \hat{C}_{2222} = -\mu$.
Replacing the fast-slow nonlinear interactions in Eq.~\eqref{eq: analytic -- mean-field} with the cubic closure, the three-dimensional model reduces to
\begin{subequations}
\label{eq: analytic -- mean-field-closure}
\begin{align}
    \dot{a}_1 &= \lambda a_1 - \omega a_2 - \mu a_1 (a_1^2 + a_2^2) \\
    \dot{a}_2 &= \omega a_1 - \lambda a_2 - \mu a_2 (a_1^2 + a_2^2),
\end{align}
\end{subequations}
identical to the real and imaginary parts of the invariant manifold reduction~\eqref{eq: analytic -- stuart-landau}.

\subsubsection{Energy transfer in the Burgers model}
\label{sec: results -- burgers}

Along with mean flow interactions, one of the primary challenges in applying model reduction to the Navier-Stokes equations is the under-resolution of the fine scales of the flow, which typically results in underestimating the energy dissipation.
In the classic spectral dynamics picture of isotropic turbulence, although the bulk of the energy is contained in the larger scales of the flow, the viscous energy dissipation scales with the square of the wavenumber, so that even very small scales play a key role in the energy balance of the flow.
The problem of truncating the cascade of energy from large to small scales thus applies equally well to ``model reduction'' in the sense of discretizing the governing equations on a mesh that cannot resolve the finest scales of the flow.

Since POD modes are ordered hierarchically by average energy content, this problem often manifests in reduced-order modeling when attempting to severely truncate a POD representation; the higher-order modes are often much more dissipative, even though they may not be of practical interest~\citep{Noack2011book}.
Further complicating this is the issue that advection-dominated flows may require many more modes for a good kinematic approximation than for a good dynamic approximation, and these additional degrees of freedom can lead to decoherence and instability in the reduced-order model~\citep{Callaham2022nonlinear}.
The approach of partitioning the system and averaging over the ``fast'' variables can address both of these issues by retaining only the most energetic degrees of freedom while accounting for the average effect of unresolved dissipation at fine scales.

To illustrate the role of nonlinear advection,~\citet{Burgers1948} and~\citet{Hopf1948} introduced simple one-dimensional toy models of turbulence, now generally studied as Burgers` equation.
Here we apply the multiscale closure approach to the Fourier-Galerkin representation of a forced, viscous Burgers` equation on a periodic domain~\citep{Noack2008jnet}:
\begin{equation}
    \label{eq: results -- burgers-eqn}
    \pdv{u}{t} + u \pdv{u}{x} = \eta \pdv[2]{u}{x} + \zeta g(x, t), \qquad x \in (0, 2\pi),
\end{equation}
where $\eta$ and $\zeta$ are the viscous and forcing amplitudes, respectively, and the forcing $g(x, t)$ is constructed to act on only the leading wavenumbers.
\citet{Majda2006nonlinearity} used a similar model to explore a similar stochastic model reduction procedure on large deterministic systems without asymptotic scale separation.

Since the domain is periodic, we forego the empirical POD basis and expand the solution in a Fourier basis
\begin{equation}
    u(x, t) = \sum_{k=-\infty}^\infty a_k(t) e^{ikx}.
\end{equation}
The standard Galerkin projection gives the coupled system of ODEs
\begin{equation}
\label{eq: results -- burgers-galerkin}
    \dot{a}_k = -\eta k^2 a_k - i \ell a_{\ell} a_{k - \ell} + \zeta \delta_{1|k|} a_k,
\end{equation}
where the forcing reduces to the Kronecker delta function $\delta_{1|k|}$ by construction.
The quadratic nonlinearity would typically result in a double sum over wavenumbers, but the orthogonality property of the Fourier basis restricts the interaction to the ``triads'' $(k, \ell, k - \ell)$.
As for the full Navier-Stokes equations, this nonlinearity satisfies the conservative property discussed in Sec.~\ref{sec: galerkin -- symmetries}.
The system is forced at large scales $|k| = 1$ and is dissipative at smaller scales $|k| \gg 1$, and the triadic interactions mediate energy transfers between these scales.
Although the closure method described in Sec.~\ref{sec: closure} was derived for real-valued systems, Eq.~\eqref{eq: results -- burgers-galerkin} could be transformed to be purely real by separating the real and imaginary parts of the complex-conjugate pairs, as was done for instance by~\citet{Noack2008jnet}.
In this case the system is simple enough that it does not matter, so we work with the complex representation for compact notation.

We can apply the results of Sec.~\ref{sec: closure} directly by rewriting the quadratic term with the sparse tensor $\mathsfbi{Q}$ defined by
\begin{equation}
    Q_{k \ell m} = \begin{cases}
    -i \ell, \qquad & \ell + m = k\\
    0,\qquad & \mathrm{otherwise}.
    \end{cases}
\end{equation}
The closure modeling is simplified by the diagonal structure of the linear term, so that the only modification to Eq.~\eqref{eq: results -- burgers-galerkin} is the cubic term $\hat{\mathsfbi{C}}$ given by Eq.~\eqref{eq: closure -- closed-tensors-cubic}.
Based on the sparse structure of $\mathsfbi{Q}$, the sum over fast scales  disappears so that
\begin{equation}
    \hat{C}_{k \ell m n} =
    \frac{1}{\nu_{k - \ell}} \begin{cases}
      -k m, \qquad & \ell + m + n = k \\
     0, \qquad & \mathrm{otherwise}.
    \end{cases}
\end{equation}
Note that these entries are also nonzero only for $k$, $\ell$, $m$, and $n$ in the slow scales and $k - \ell$ in the fast scales.
For example, if we model only the forced scales $|k|=1$ as slow variables, the only nonvanishing contributions to $\hat{\mathsfbi{C}}$ are from the wavenumber sets $(k, \ell, m, n) = (1, -1, 1, 1) $ and $(-1, 1, -1, -1)$.
Since the slow quadratic tensor $\hat{\mathsfbi{Q}}$ is also zero for $|k|=1$, the closed ODE system simplifies to the one-dimensional Stuart-Landau equation
\begin{equation}
\label{eq: results -- burgers-closed}
    \dot{a}_1 = (\zeta - \eta) a_1 - \nu_2^{-1} a_1 |a_1|^2.
\end{equation}

The cubic term in Eq.~\eqref{eq: results -- burgers-closed} captures the leading-order influence of the energy transfers to smaller scales.
This effect is naturally stabilizing; as in the physical mechanism, if the slow variables are excited, the amplitude term $|a_1|^2$ representing the unresolved scales increases and the cubic damping term removes energy from the resolved scales.

In this case the multiscale closure directly accounts for only the next set of wavenumbers $|k|=2$ because the purely triadic structure of the quadratic interactions does not admit contributions from any smaller scales.
The damping constant $\nu_2$ must therefore be selected so that the explicitly modeled transfers from $|k|=1$ to $|k|=2$ are representative of the entire cascade.
This can be accomplished with the energy balance method described in Sec.~\ref{sec: closure}, for example.

While traditional closure models for Galerkin systems approximate the dissipation at higher orders with a linear linear eddy-viscosity-type closure~\citep{Aubry1988jfm, Rempfer1994}, the mechanism of energy transfer to smaller scales is fundamentally nonlinear in nature.
As clearly illustrated by this model problem, the multiscale closure model preserves the nonlinearity of the energy transfer in its approximation of the effects of the unresolved small scales.
This is consistent with the findings of~\citet{Osth2014jfm}, that a nonlinear term modeling subscale turbulence can outperform the linear eddy viscosity closure in POD-Galerkin models.

\subsection{Vortex shedding in the cylinder wake}
\label{sec: results -- cylinder}

\begin{figure}
\centering
\begin{overpic}[width=0.8\linewidth]{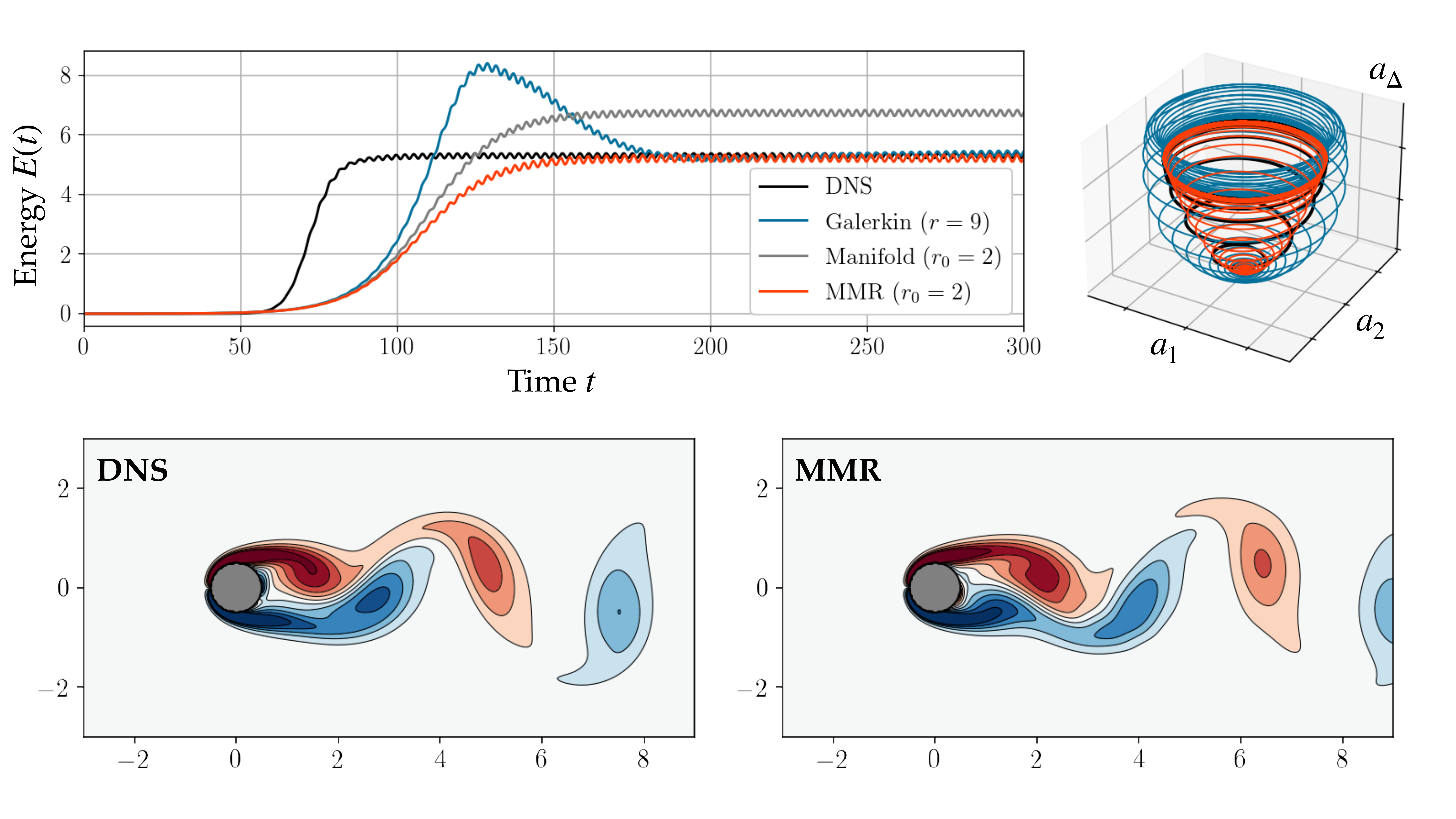}
\put(2,52.5){a)}
\put(73,52.5){b)}
\put(2,25){c)}
\end{overpic}
\vspace{-.25in}
\caption{
{\small
\textbf{Reduced-order models of the cylinder wake.}
The standard 9-mode Galerkin model accurately estimates the stable limit cycle (a), but the transient dynamics deviate from the slow manifold, leading to an energy overshoot (b).
Both the two-dimensional invariant manifold reduction~\citep{Noack2003jfm} and MMR closure models prevent this by eliminating the rapidly-equilibrating variable associated with mean flow deformation, but MMR more accurately estimates the limit cycle amplitude.
The full flow field can be reconstructed by approximating the harmonics with polynomial regression (c).
}
}
\label{fig: results -- cyl}
\end{figure}

In this section we revisit the canonical two-dimensional flow past a circular cylinder to show that the MMR closure can naturally resolve the stabilizing effects, both of mean flow deformation and of energy transfer to higher-order modes, without assuming proximity to a bifurcation.

In keeping with previous work demonstrating that the cylinder wake can be modeled accurately with as few as two degrees of freedom, we reduce the 9-mode POD-Galerkin system to a 2-mode generalized Stuart-Landau model~\eqref{eq: closure -- stuart-landau} with the closure model~\eqref{eq: closure -- closed-tensors} and damping derived from the energy-balance relation~\eqref{eq: closure -- energy-balance}.
In this case, the elimination of the higher-order modes does not require sacrificing kinematic resolution, since the unresolved coefficients can be approximated as sparse polynomial functions of the resolved variables~\citep{Loiseau2018, Callaham2022nonlinear}.

Figure~\ref{fig: results -- cyl}a compares the energy of the first mode pair, $E(t) = a_1^2(t) + a_2^2(t)$, predicted by the POD-Galerkin model and the 2-mode closure model to DNS of the transient flow initialized from the unstable steady state and perturbed by the least-stable eigenmode.
Also included for comparison is the invariant manifold model introduced by~\citet{Noack2003jfm} and described in Sec.~\ref{sec: closure -- invariant}.
Although the Galerkin model is stable and accurately predicts the amplitude of the vortex shedding, it exhibits a transient energy overshoot before settling onto the limit cycle.
Figure~\ref{fig: results -- cyl}b shows the reason for this; whereas the DNS is restricted to an approximately parabolic invariant manifold, the finite relaxation time in the shift mode dynamics results in the energy of the leading modes continuing its exponential growth for longer before the energy can be absorbed into the mean flow deformation.

In contrast, the cubic term in the MMR and invariant manifold models approximates the relaxation of the shift mode dynamics as instantaneous, pinning its amplitude to that of the primary mode pair.
While the structure and interpretation of the two models is the same, there are quantitative differences in their accuracy.
In particular, the manifold model slightly overestimates the energy of the limit cycle, with mode amplitudes larger than the DNS by $\sim 13\%$, while the multiscale closure accurately predicts the limit cycle amplitude.
This discrepancy can most likely be attributed to the energy balance procedure used to estimate the subscale damping $\bm \nu$ in the MMR model, which is not used in the invariant manifold model.

Finally, the full flow field can also be reconstructed from the two degree-of-freedom multiscale model by means of the sparse polynomial approximation to the nonlinear correlations in the higher-order modes (Fig.~\ref{fig: results -- cyl}c). 
The flow field predicted by the model gives a good approximation of the DNS solution, although over long times the phase of the vortex shedding tends to drift from that of the high-dimensional solution.
By systematically averaging over the fast variables, the multiscale closure method is thus able to reduce the dimensionality of the model while also improving its fidelity.

\subsection{Chaotic lid-driven cavity flow}
\label{sec: results -- cav}

The Stuart-Landau equations derived for the analytic problems in Sec~\ref{sec: results -- toy-problems} and the cylinder wake in Sec.~\ref{sec: results -- cylinder} are examples of relatively simple nonlinear oscillators with either a stable fixed point or periodic limit cycle as an equilibrium.
Here we show that the MMR closure approach can also be applied to systems with more complex dynamics, as illustrated on the chaotic lid-driven cavity flow introduced in Sec.~\ref{sec: numerical -- cav}.

\begin{figure}
\centering
\includegraphics[width=0.8\textwidth]{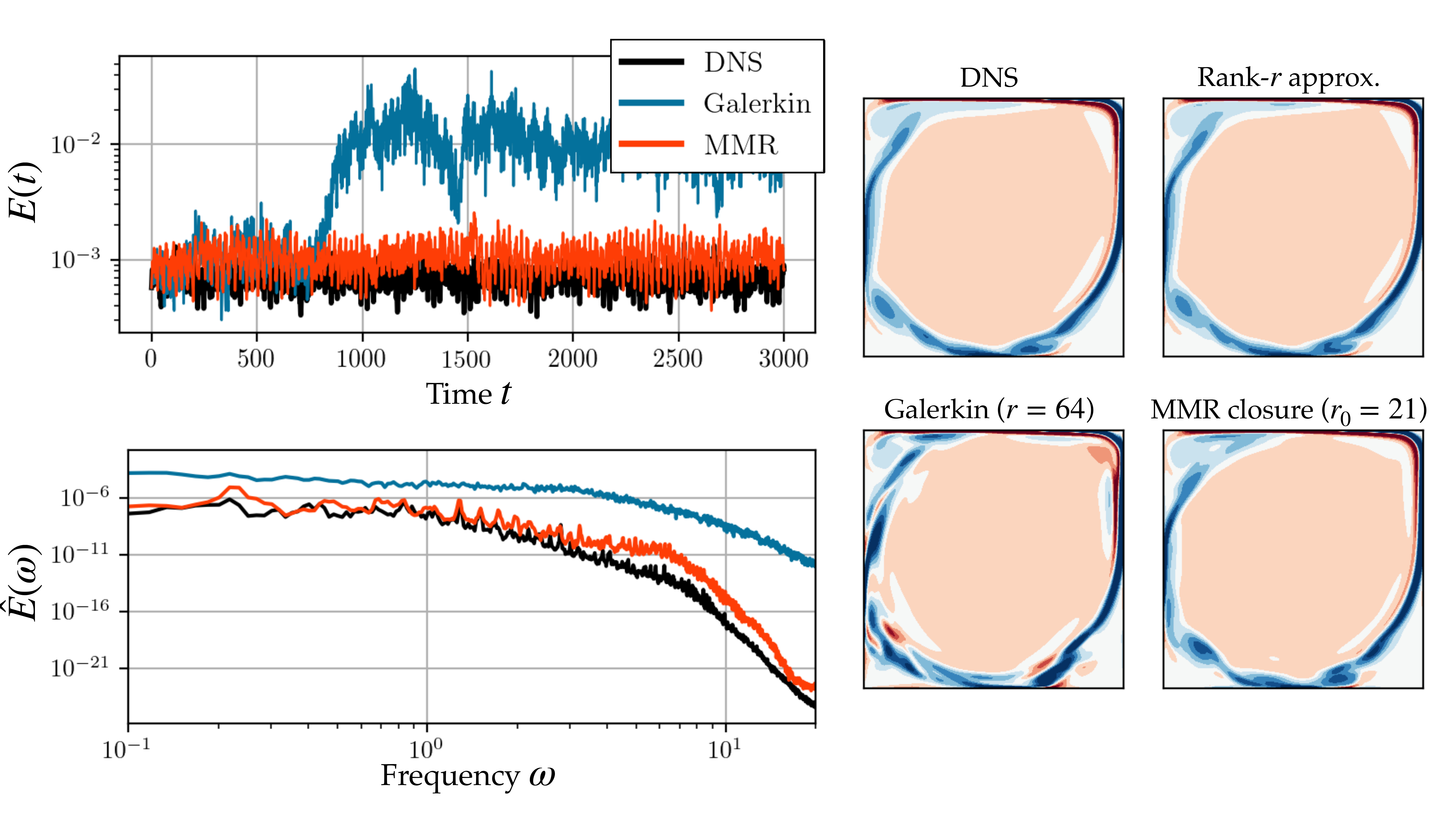}
\vspace{-.25in}
\caption{{\small
\textbf{Reduced-order models of the lid-driven cavity flow.}
The Galerkin system eventually equilibrates at much higher energy levels, while the MMR closure model produces stable predictions that closely match the power spectrum of the chaotic flow (left)
The reconstructed fields from the MMR model are also more physically consistent with the DNS than the Galerkin model (right).
}}
\label{fig: results -- cav}
\end{figure}

First, we construct a standard POD-Galerkin model with $r=64$ modes, sufficient to capture $\sim 99.8\%$ of the fluctuation kinetic energy.
The Galerkin system is chaotic and energetically stable. 
Despite its precise kinematic resolution, the Galerkin model is inaccurate, overestimating the kinetic energy by roughly one order of magnitude, as shown by the blue traces in Fig.~\ref{fig: results -- cav}.

Reducing the model to $r_0 = 21$ degrees of freedom via a multiscale closure stabilizes it with an equilibrium energy that approximately matches the power spectrum of the DNS (Fig.~\ref{fig: results -- cav}, orange).
The MMR model still retains enough modes for a $\sim 98\%$ accurate POD reconstruction.
The predicted flow fields are coherent and qualitatively consistent with the DNS fields, even though they do not match in every detail due to the chaotic nature of the flow.

Varying the dimension $r_0$ of the approximate slow variables results in models that are energetically stable, but not necessarily chaotic.
Instead, many such models eventually settle onto post-transient dynamics that are periodic or quasi-periodic.
The persistence of the dominant frequencies suggests that even for models with relatively many degrees of freedom ($r_0>2$), the generalized Stuart-Landau equations produced by the MMR closure scheme can be viewed as coupled nonlinear oscillators, especially when the spatial components of the POD modes suggest that they may be grouped in pairs with similar temporal frequency content.

\subsection{Convective instability and vortex pairing in a mixing layer}
\label{sec: results -- mixing-layer}

Advection-dominated flows have historically posed a challenge for reduced-order models, primarily because the global mode ansatz of POD is a poor representation of traveling waves.
As a final numerical example, we apply the MMR closure to model the mixing layer introduced in Sec.~\ref{sec: numerical -- mix}, a canonical example of convective instability in free shear flows.

In one sense, the dynamics of the mixing layer are relatively simple.
The flow close to the inlet is dominated by the shear layer instability and subsequent vortex roll-up.
These vortices are advected downstream at approximately the midline flow velocity, until the secondary vortex pairing begins at the subharmonic of the primary instability frequency.
However, since this behavior unfolds as the disturbances are carried downstream, a principal challenge for low-dimensional models based on global POD modes is to preserve phase coherence between the harmonics and subharmonics, so that the reconstructed flow fields consist of approximately discrete vortices undergoing the vortex pairing.

As discussed in Sec.~\ref{sec: numerical -- mix}, both the POD mode basis and the difficulty of modeling this flow depend on the streamwise extent of the domain.
Since the subharmonic vortex pairing is energetically dominant downstream, the leading POD modes computed from a longer domain will be related to vortex pairing, even though the primary shear layer instability drives the dynamics from a physically causal perspective.
To illustrate this effect, we consider two domain lengths: one that is dominated by the primary instability and vortex roll-up, and the full modeling domain with vortex pairing.

\begin{figure}
\centering
\begin{overpic}[width=\linewidth]{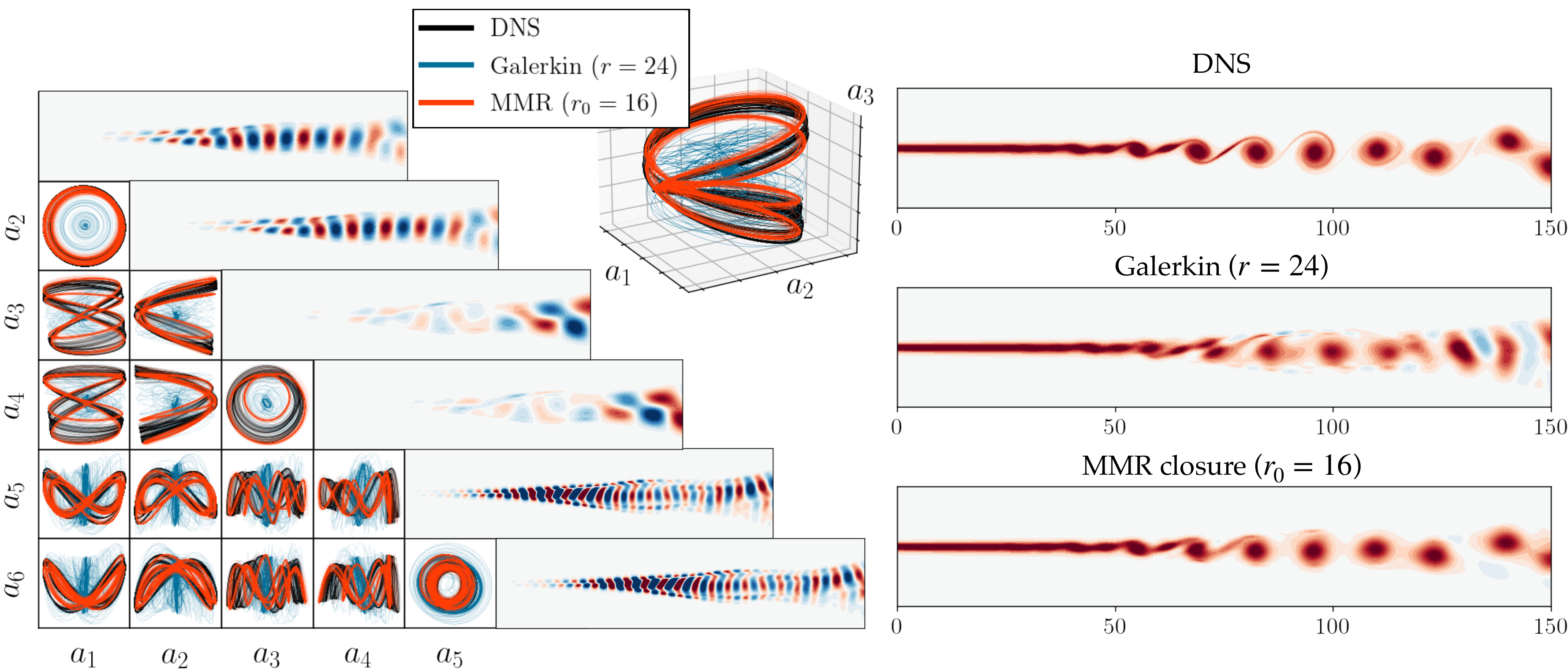}
\end{overpic}
\caption{
{\small
\textbf{Models of the mixing layer on a short domain.}
The phase portraits of the mode pairs (left) show phase-locking between the fundamental mode pair $(a_1, a_2)$, the subharmonic vortex pairing $(a_3, a_4)$, and the first harmonic $(a_5, a_6)$.
These phase relationships are preserved by the MMR closure, resulting in physically consistent flow field estimates (right).
} }
\label{fig: results -- mix-short-results}
\end{figure}

\begin{figure}
\vspace{-.05in}
\centering
\begin{overpic}[width=0.9\linewidth]{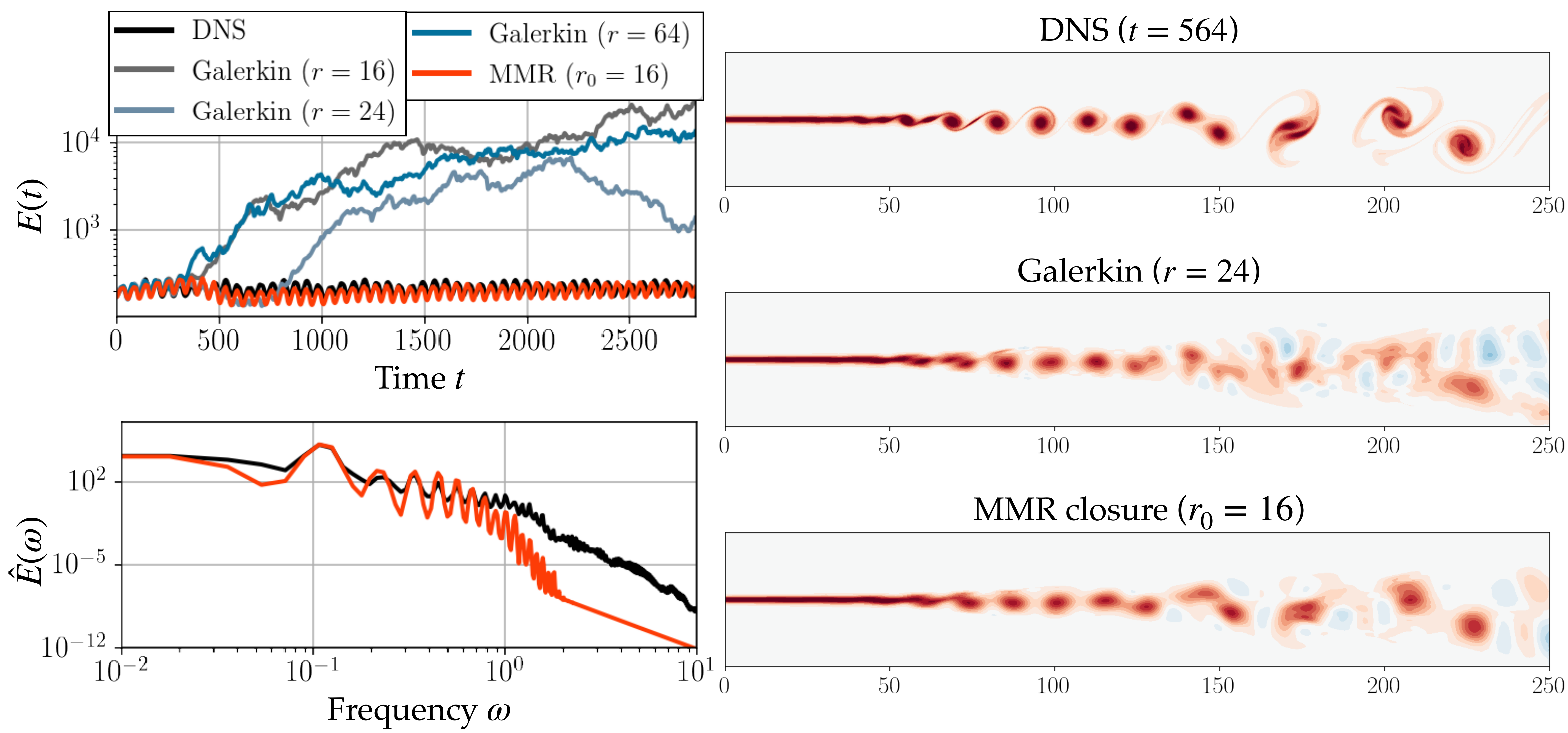}
\end{overpic}
\vspace{-.1in}
\caption{
{\small
\textbf{Mixing layer models on the long domain.}
While standard POD-Galerkin models are energetically unstable until at least $r=64$, the multiscale closure stabilizes the model with only 16 variables (left).
The model also remains coherent on long integration times (see also Fig.~\ref{fig: results -- mix-phase}), producing flow field predictions that are consistent with the large-scale structure of the DNS (right).
} }
\label{fig: results -- mix-results}
\end{figure}

In this case the POD-Galerkin models do not necessarily improve with an increasing number of modes. 
For both domains, the best dynamical approximation is found for models constructed from $r=24$ modes.
On the short domain ($L_x = 150$), this model is energetically stable but eventually approaches an unphysical fixed point (Fig.~\ref{fig: results -- mix-short-results}). 
On the longer domain ($L_x = 250$), the model initially follows the DNS trajectory before eventually becoming unstable (Fig.~\ref{fig: results -- mix-results}).
In either case, increasing the dimension of the POD-Galerkin model resulted in worse models.
We then reduce the $24$-dimensional Galerkin systems to $r_0=16$-dimensional generalized Stuart-Landau models with the MMR closure.

Figure~\ref{fig: results -- mix-short-results} shows the comparison of the closed model to the DNS and Galerkin systems for the short domain.
The MMR model closely matches the phase portraits of the DNS, indicating that it preserves the phase relationships between the various modes, including the early subharmonic vortex pairing $(a_3, a_4)$ and higher harmonics, e.g. $(a_5, a_6)$.
This remains true even when integrated to very long times (the final integration time is $t=2820$ in this case, as with the DNS described in Sec.~\ref{sec: numerical -- mix}).
As a result, the approximated fields are highly coherent, with the same pattern of vortex roll-up, advection, and pairing as seen in the DNS (Fig.~\ref{fig: results -- mix-short-results}, right).

\begin{figure}
\centering
\begin{overpic}[width=\linewidth]{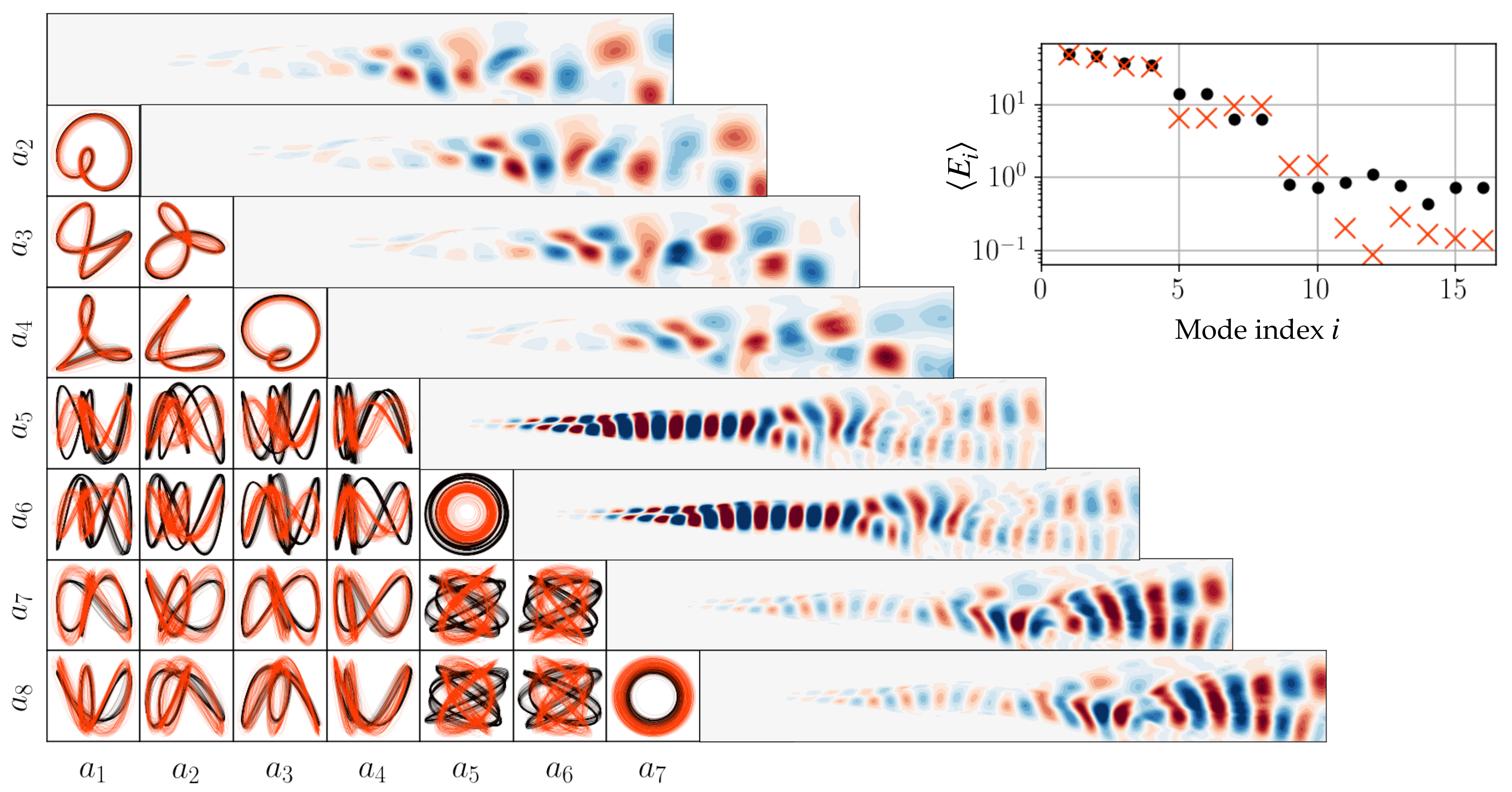}
\end{overpic}
\caption{
{\small
\textbf{Phase portraits of the mixing layer model on the long domain.}
The MMR closure (orange) closely matches the DNS trajectories (black) for the most energetic modes, even though the structure is more complicated than the Lissajous-type harmonics on the short domain (Fig.~\ref{fig: results -- mix-short-results}).
This figure does not show a Galerkin model for comparison because they are energetically unstable (Fig.~\ref{fig: results -- mix-results}).
} }
\label{fig: results -- mix-phase}
\end{figure}

On the longer domain, the multiscale closure also stabilizes the Galerkin model.  
Figure~\ref{fig: results -- mix-results} compares the true energy content of the POD coefficients to that predicted by the model, both in the time and frequency domain.
Although the evolution of the energy matches at the dominant frequencies, the MMR model does not capture the weakly chaotic dynamics exhibited by the DNS as a result of slight irregularity in the vortex pairing.
Instead, after a brief transient the solution becomes approximately periodic with phase locking between the modes (Fig.~\ref{fig: results -- mix-phase}).
As with the short domain, this synchronization ensures that the predicted flow fields remain coherent, even when integrated for long times (Fig.~\ref{fig: results -- mix-results}, right).

As might be anticipated based on the more complex dynamics on the longer domain, the fields predicted by the MMR model do not match the DNS quite as closely as on the shorter domain, particularly closer to the inlet.
This is likely because fine-scale resolution of the upstream region of the flow requires higher harmonics of the primary instability that do not appear in the POD modes of the longer domain until much higher-order mode number.
Higher-resolution predictions might therefore be constructed by applying nonlinear correlations analysis and polynomial regression to selected higher-order modes, as in the cylinder reconstruction in Sec.~\ref{sec: results -- cylinder}; see also~\citet{Loiseau2018} and~\citet{Callaham2022nonlinear}.
Still, the dominant vortices are still evident in the prediction of the low-dimensional model, even in the downstream pairing region.

On both domains, the synchronization and long-time phase coherence between the modes further reinforces the picture outlined in Sec.~\ref{sec: results -- cav} of the closure model as consisting of coupled nonlinear oscillators.
Although the description of this flow in terms of spatially fixed global modes is not an ideal representation of the advection-driven dynamics, the proposed MMR closure method is able to successfully stabilize the POD-Galerkin models and accurately capture the phase relationships necessary for accurate predictions of the full flow fields.

%% file: S7_Discussion.tex
\section{Discussion}
\label{sec: discussion}

In this work, we have developed a multiscale model reduction (MMR) approach to improve linear-quadratic dynamical systems, derived from POD-Galerkin projection of the Navier-Stokes equations, with the addition of systematically computed cubic closure terms.  
These cubic closure terms are derived through an adaptation and application of a multiscale stochastic averaging method that originated in singular perturbation theories of Markov processes~\citep{Kurtz1973, Papanicolaou1976, Majda2001cpam, EWeinan2011book, PavliotisStuart2012}. 
Whereas the standard truncation of the Galerkin system disregards the influence of unresolved variables, the proposed multiscale model reduction method, accounts for their effect in an average sense through averaging via the stochastic Koopman operator.
In particular, this approach is able to model nonlinear interactions between resolved and unresolved variables, capturing key mechanisms such as mean flow deformation and the energy cascade.
The closed model includes cubic terms, taking the form of generalized Stuart-Landau equations that often act as coupled nonlinear oscillators.

Since the derivation of the MMR closure method relies on an asymptotic timescale separation between resolved and subscale variables that is generally absent in fluid flows, we do not attempt to prove the general validity of this approach when the flow is not close to a bifurcation.
Instead, we have demonstrated by extensive numerical examples that the method dramatically improves the stability and accuracy of low-dimensional models of unsteady fluid flow, compared to the standard POD-Galerkin models.
After a comparison to the well-known benchmark problem of vortex shedding in a cylinder wake, we have shown that the MMR method can reproduce chaotic dynamics in lid-driven cavity flow.
Finally, we developed models of an incompressible mixing layer and showed that the coupled-oscillator MMR closures not only stabilize the models, but preserve phase relationships between the modes, which is critical for physically consistent predictions of advection-driven flow.

These results raise several possibilities for interesting future work.
For example, the mixing layer is an ``amplifier'' flow characterized by convective instability and transient energy growth.
It is therefore highly sensitive to the nature of the upstream flow; models developed for a specific inlet forcing may not generalize.
Further consideration of this issue could be the foundation of a multiscale approach to input/output models in the resolvent framework~\citep{McKeon2010jfm, Sharma2013jfm, zare2017colour, Padovan2020jfm, Pickering2021jfm}.
In this work we neglect the time-varying boundary condition in the region upstream of the modeling domain, but an input/output model could more properly treat this explicitly as forcing, as is done for instance in control-oriented reduced-order models~\citep{Barbagallo2009jfm, Sipp2010amr}.
Recent work on trajectory-based optimal oblique projections has also shown promise for capturing dynamically important, but typically low-energy, modes in an input-output framework for model reduction of open shear flows~\citep{Otto2022}.

As discussed for the chaotic lid-driven cavity flow in Sec.~\ref{sec: results -- cav}, we have observed that the MMR models often tend towards phased-locked periodic or quasiperiodic solutions rather than chaos, consistent with the idea of a system of weakly coupled oscillators.
However, we expect that in the development of models of chaotic or turbulent flows it may be helpful to re-introduce some degree of stochasticity to better match the flow statistics.
Fortunately, this is straightforward in the multiscale modeling framework~\citep{EWeinan2011book, PavliotisStuart2012}.
For instance, the homogenization method~\citep{Majda2001cpam} begins from slightly different scaling assumptions and results in a closed model in which the fast-scale stochastic forcing appears as a new diffusion term.
A careful investigation of different scale separation assumptions and stochastic approximations could be valuable in extending this framework to turbulent flows.

In this work we have only considered linear-quadratic dynamics derived from POD-Galerkin projection.
This has the advantage of being based on the governing equations, but recent work has shown that data-driven model discovery can be a powerful alternative~\citep{Brunton2016pnas, Loiseau2017jfm, Loiseau2018, Loiseau2018jfm, Qian2020, Rubini2020jfm, Deng2020jfm, Callaham2022nonlinear}.
Data-driven modeling is especially useful when the governing equations are unknown or projection-based modeling is not as straightforward as for incompressible fluid flow~\citep{Qian2020, Loiseau2020tcfd, Guan2021electro}.
In a sense, data-driven modeling might circumvent the need for closure modeling by fitting directly to the time series, including leveraging the intuition that cubic terms will appear in the effective dynamics~\citep{Loiseau2017jfm}.
However, the linear-quadratic system has certain symmetries and conservation properties that can be enforced in constrained regression to a quadratic model~\citep{Schlegel2015jfm, Loiseau2017jfm, Kaptanoglu2021, Kaptanoglu2021pre}, while extending these properties to cubic nonlinearity is not straightforward.
An interesting future direction could be exploring a two-step process, first identifying a quadratic model that satisfies the constraints, and then applying the MMR closure to reduce it to a lower-dimensional cubic model.
Since the empirical models are often sparse, this could potentially result in models with improved stability and scaling compared to the projection-based approach.
Alternatively, a more flexible deep-learning model could be used to replace the proposed diagonal drift approximation, for instance with a modified variational autoencoder~\citep{Kingma2013vae} that parameterizes the distribution of subscale variables conditioned on the slow variables.

It will also be interesting to extend this analysis to systems where the leading-order cubic terms tend to \emph{destabilize} the system, such as the subcritical bifurcation in plane Poiseuille flow~\citep{Stewartson1971}.
This is an important scenario in the transition to turbulence, particularly for shear flows in which non-normal energy amplification can activate the nonlinearity even when the flow is globally stable.
Higher-order effective nonlinearity could potentially be incorporated into the framework via state-dependent diffusion or an alternative fast drift model, for instance.

Finally, the analytic elimination of small scales in the Fourier-Galerkin model of Burgers' equation raises the possibility of a subscale LES closure model based on multiscale analysis.
This might take the form of a multiple-scale expansion in both space and time or of a filter designed for the isotropic scales of turbulence, for instance.
In either case, the Stuart-Landau form of the MMR closure model suggests that a spatioteporal analysis would be akin to a Ginzburg-Landau theory of variational multiscale turbulence modeling.

Beyond the context of POD-Galerkin reduced-order modeling, the multiscale model reduction framework represents a general theory of effective cubic nonlinearity arising in amplitude equations for global modes in quadratically nonlinear fluid flow.
The system of generalized Stuart-Landau equations produced by the closure model presents a picture of the flow as a set of coupled nonlinear oscillators describing the evolution of mode pairs.
Close to the threshold of instability the MMR closure is consistent with a weakly nonlinear analysis, but it does not rely on proximity to a bifurcation.
The multiscale modeling methodology is therefore an alternative to standard asymptotic expansions that may have implications for a variety of theoretical and numerical approaches to modeling the large, slow scales of unsteady fluid flows.

%% file: A1_Stability.tex
\section{Stability analysis of the mixing layer inlet profile}
\label{app: stability}

\begin{figure}
\centering
\includegraphics[width=\textwidth]{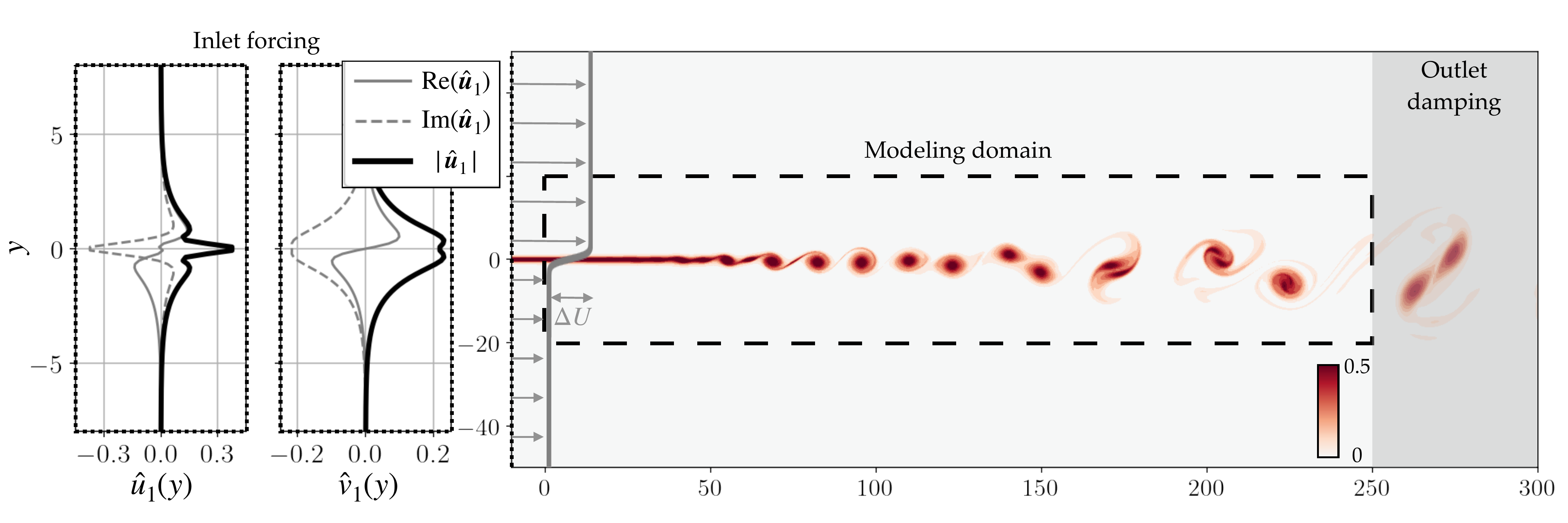}
\caption{
{\small
\textbf{An incompressible mixing layer at $\Rey = 500$.}
The hyperbolic tangent base flow is convectively unstable, amplifying small perturbations as they are carried downstream.
We force the flow at the inlet with eigenfunctions of inviscid flow equations linearized around the base flow (left).
The central part of the domain is used for further modeling to avoid boundary effects, while the downstream extent $x \in (250, 300)$ is strongly damped to prevent numerical instability at the outlet.
}}
\label{fig: numerical -- mix-config}
\end{figure}

This appendix describes the eigenfunction forcing applied at the inlet of the mixing layer configuration described in Sec.~\ref{sec: numerical -- mix}.
The flow acts an amplifier~\citep{Huerre1985, Huerre1990arfm, Chomaz2005arfm}, so the downstream flow is highly sensitive to the inlet conditions; eigenfunction forcing accentuates the natural dynamics of the flow.

Since the dominant Kelvin-Helmholtz instability is inviscid~\citep{Michalke1964}, we derive the forcing from a temporal stability analysis of the inviscid equations linearized about the parallel hyperbolic tangent inlet profile $\bm u(\bm x) = (U(y), 0)$.
Defining a perturbation streamfunction $\psi(\bm x, t)$ with real wavenumber $\alpha$ and complex velocity $c = c_r + i c_i$:
\begin{equation}
    \psi(\bm x, t) = \Real \left\{ \hat{\psi}(y) e^{i\alpha(x - ct)} \right\},
\end{equation}
the evolution of the perturbation is given by the Rayleigh equation
\begin{equation}
\label{eq: numerical -- rayleigh}
    \left( U - c \right) \left( \dv[2]{y} - \alpha^2 \right) \hat{\psi} - U'' \hat{\psi} = 0.
\end{equation}
For each wavenumber $\alpha$, this can be written as a generalized eigenvalue problem
\begin{equation}
    \mathbfcal{A}(\alpha) \hat{\psi}_j(y; \alpha) = c^{(j)} \mathbfcal{B}(\alpha) \hat{\psi}_j(y; \alpha) 
\end{equation}
for the $j^\mathrm{th}$ velocity $c$ and perturbation streamfunction $\hat{\psi}$, where we use the convention of sorting by decreasing $\Imag\{c^{(j)}\}$.

The base profile is odd, so the real phase velocity is the midline value $c_r = \bar{U}$~\citep{Michalke1964}.
With $\bar{U} = 1$, perturbations oscillate with frequency $\omega =\alpha$ and growth rate $\sigma(\alpha) = \alpha c_i(\alpha)$.
For further details on linear stability analysis of shear flows, see for example~\citet{Huerre1985, Schmid2012book}.

We choose a spectral discretization of Eq.~\eqref{eq: numerical -- rayleigh} using a Hermite collocation scheme from a Python implementation of the \texttt{dmsuite} library~\citep{dmsuite}.
We find the least stable wavenumber $\alpha^*$ by maximizing the growth rate $\sigma$ (i.e. minimizing -$\alpha c_i^{(1)}$) using an implementation of the BFGS algorithm available in \texttt{scipy}.
Since this departs from the standard shooting method, we validate it against~\citet{Michalke1964}, who uses the profile ${U(y) = 0.5( 1 + \tanh y )}$ based on nondimensionalizing with the shear layer thickness $\delta_\omega/2$ rather than vorticity thickness.
With that profile we find a least-stable wavenumber $\alpha^* = 0.4449$ and growth rate $\sigma = 0.0949$ compared to the reported results of $\alpha^* = 0.4446$ and $\sigma = 0.0949$.
For the present base profile~\eqref{eq: numerical -- tanh} we find $\alpha^* = 0.8912$ and $\sigma = 0.2129$.

Once the wavenumber $\alpha^*$ and eigenfunction $\hat{\psi}_1$ corresponding to the maximum growth rate is identified, the inlet velocity perturbation can be computed from
\begin{equation}
    \hat{u}_1(y; \alpha^*) = \hat{\psi}_1'(y; \alpha^*) , \hspace{2cm} \hat{v}_1(y; \alpha^*)  = i \alpha^* \hat{\psi}_1(y; \alpha^*) .
\end{equation}
The optimal perturbations are shown in Fig.~\ref{fig: numerical -- mix-config}.
Following~\citet{Colonius1997}, we also apply forcing at the first three subharmonics of the least-stable mode computed from solving~\eqref{eq: numerical -- rayleigh} at $\alpha^*/2$, $\alpha^*/4$, and $\alpha^*/8$, rescaling each so that the maximum value of $\hat{u}_1$ is $10^{-3} \Delta U$.
We perturb the inlet profile by $\Real\{ \hat{\bm u}_1(y, \alpha_n) e^{i \omega_n t} \}$ for $n=1, 2, 4, 8$ and $\alpha_n = \omega_n = \alpha^*/n$ and use a simulation time step of $\Delta t = 0.00705 = 10^{-3} \times 2 \pi / \alpha^*$, so that the sampling rate of the simulation is commensurate with the forcing frequency.

%% file: A2_Truncation.tex
\section{Variable truncation of the mixing layer model}
\label{app: truncation}

One of the limitations of projection-based reduced-order modeling for systems without an explicit scale separation is a lack of principled criteria for selecting the truncation rank, or dimension of the projection subspace.
A common approach is to use \emph{ad hoc} criteria such as the number of modes required to resolve some fraction of the fluctuation kinetic energy (e.g. 90\% or 99\%) on average.
To some extent this simply replaces the arbitrary choice of rank truncation with an equally arbitrary choice of approximation accuracy.
This is especially the case when the dynamic approximation accuracy of the reduced-order model does not monotonically improve with increasing kinematic approximation accuracy (as can be seen in Fig.~\ref{fig: results -- mix-results}, for instance).

Unfortunately, this problem is compounded in the secondary reduction step of the MMR method, since one must select both the rank $r$ of the original POD-Galerkin system and the rank $r_0$ of the final MMR system.
Our approach in this work was to follow the heuristic of simulating the POD-Galerkin system at various values of $r$ and choosing that which performs best or remains stable for longest (e.g. $r=24$ for the mixing layer model in Fig.~\ref{fig: results -- mix-results}).
The final MMR models then typically behave similarly for a range of final ranks $r_0$; we report results for those that best reproduce the dynamics of the fluid flow.

\begin{figure}
\centering
\includegraphics[width=0.9\textwidth]{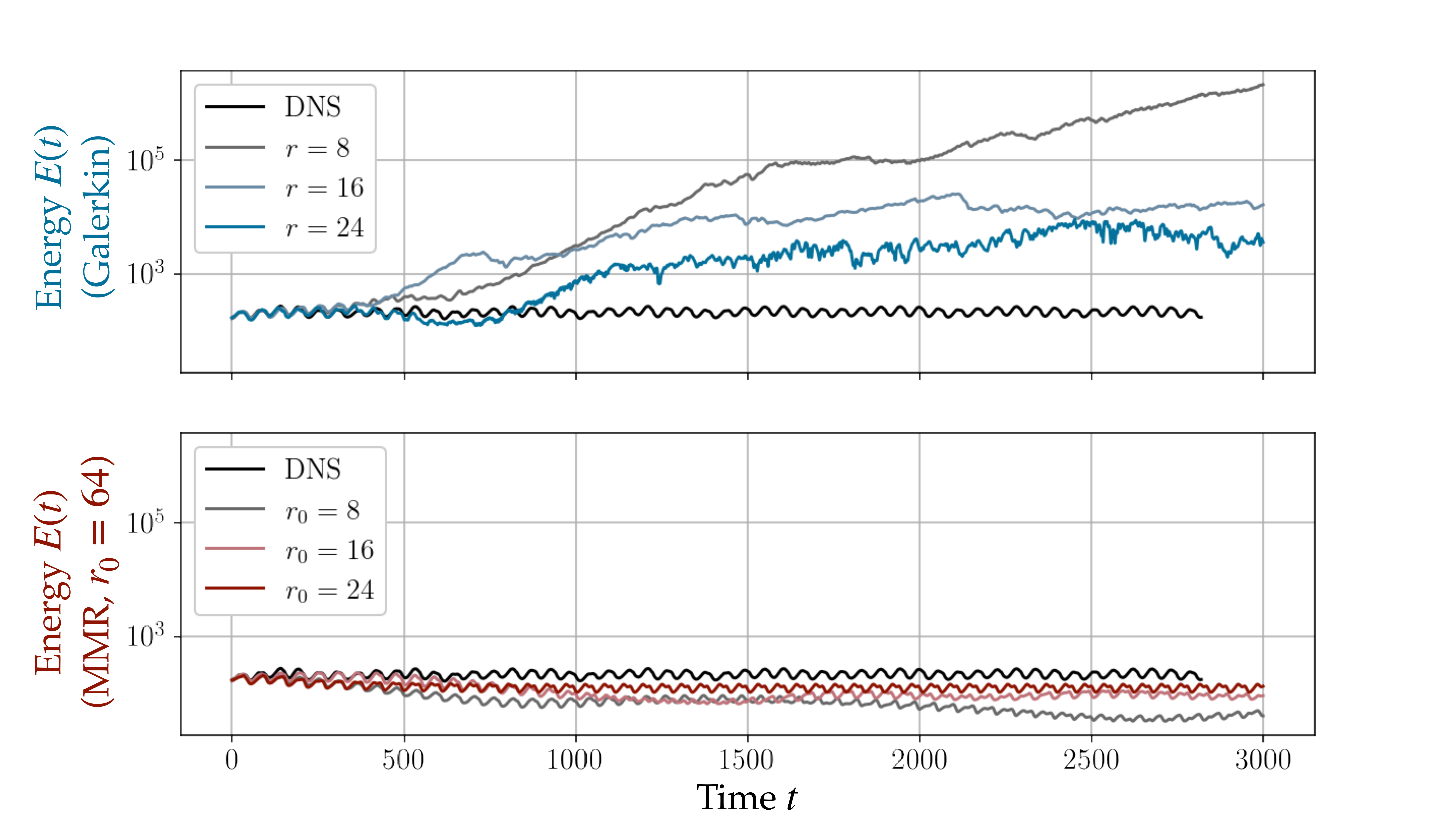}
\caption{
{\small
\textbf{Alternative truncations of the mixing layer model in Sec.~\ref{sec: results -- mixing-layer}.}
Rather than the ``optimal'' value of $r=24$, the MMR models shown here are constructed from Galerkin systems of rank $r=64$.
While the POD-Galerkin models are inaccurate and energetically unstable for any rank truncation, while the MMR models are less sensitive to the specific values of rank truncation.
}
}
\label{fig: truncation}
\end{figure}

However, in order to show that the method is relatively insensitive to the specific choices of rank truncation, here we repeat the model reduction for the mixing layer on the longest domain shown in Fig.~\ref{fig: results -- mix-results}.
These models begin with $r=64$ rather than the ``optimal'' POD-Galerkin rank of $r=24$.
Although less accurate than the model reported in Sec.~\ref{sec: results -- mixing-layer}, the resulting models are stable for all values $r_0$ of the final rank and significantly outperform any POD-Galerkin model.

\begin{figure}
\centering
\includegraphics[width=0.85\textwidth]{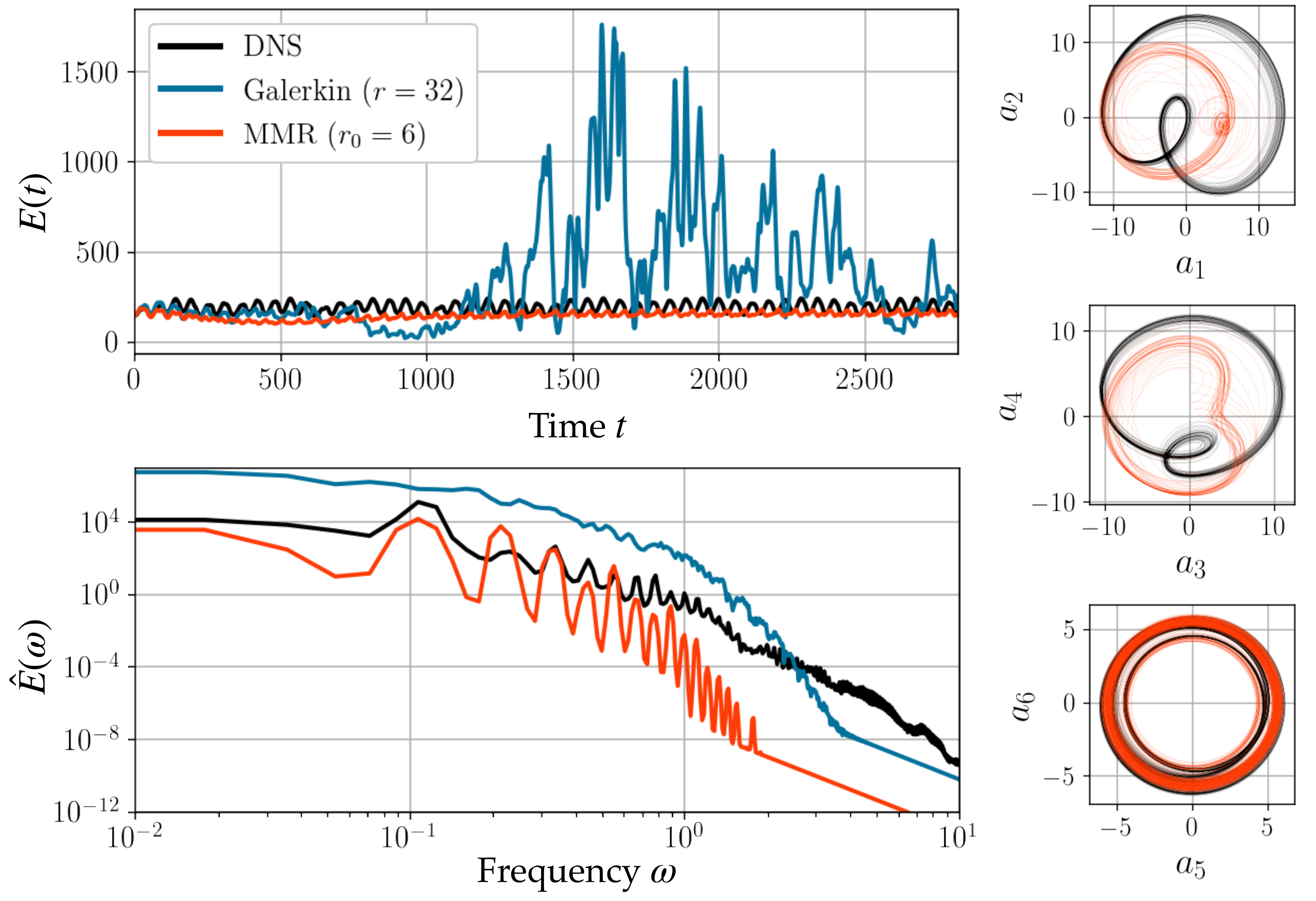}
\caption{
{\small
\textbf{Truncation of the mixing layer in Sec.~\ref{sec: results -- mixing-layer} based on the singular value spectrum.}
In this case the initial truncation level $r=32$ is selected based on a residual of 1\%, while the MMR rank $r_0=6$ is based on the ``knee'' criterion (see text).
As with the models shown in Fig.~\ref{fig: truncation}, the resulting model is stable and roughly matches the true dynamics, but is not as accurate as the model reported in Sec.~\ref{sec: results -- mixing-layer}.
}
}
\label{fig: opt-truncation}
\end{figure}

As a more principled example, Fig.~\ref{fig: opt-truncation} shows a model constructed by choosing a kinematic reconstruction accuracy threshold of 99\% for the POD-Galerkin rank $r$.
The MMR rank $r_0$ is then selected with the ``knee'' criterion, which selects a truncation level based on the inflection point of the singular value distribution.
Again, the model is less accurate than the results shown in in Sec.~\ref{sec: results -- mixing-layer}, but significantly more stable and accurate than any Galerkin system.
Thus, while a principled approach to rank selection is an open problem for both POD-Galerkin modeling and the proposed multiscale reduction, the latter appears generally less sensitive to this choice.

%% file: A3_Closures.tex
\section{Nonlinear eddy viscosity model}
\label{app: closure}

Multiscale model reduction is similar in spirit to the finite-time thermodynamics (FTT) closure framework introduced by~\citet{Noack2008jnet}, but notably different in its particulars, as shown in Sec.~\ref{sec: closure -- ftt}.
One drawback to FTT closure modeling is that it increases the state dimension by a factor of 2, since the fluctuation energies $\{E_i\}$ have to be simulated along with the state variables.
Alternatively,~\citet{Schlegel2009},~\citet{Noack2011book}, and~\citet{Cordier2013} suggested a nonlinear eddy viscosity-type closure model based on the structure of the FTT formalism without introducing additional degrees of freedom.
For a truncated Galerkin system
\begin{equation}
    \dot{x}_i = f_i(\bm x) \equiv \sum_{j=1}^r L^B_{ij} x_j + \Rey^{-1} \sum_{j=1}^r L^V_{ij} x_j + \sum_{j, k =1}^r Q_{ijk} x_j x_k,
\end{equation}
where $\bm{L}^B$ and $\bm{L}^V$ are the parts of the linear term corresponding to base (or mean) flow advection and viscosity, respectively,
the FTT closure is based on the total mean fluctuation energy $K_\infty = \sum_{i=1}^r \overline{x_i^2} / 2$ and instantaneous fluctuation energy $K(t) = \sum_{i=1}^r x_i^2(t) / 2$.
The nonlinear FTT closure takes the form
\begin{equation}
\label{eq: ftt -- closed-galerkin}
\dot{x}_i = f_i(\bm x) + \nu_T \sqrt{\frac{K(t)}{K_\infty}} L^V_{ii} x_i.
\end{equation}

\begin{figure}
\centering
\begin{overpic}[width=0.9\linewidth]{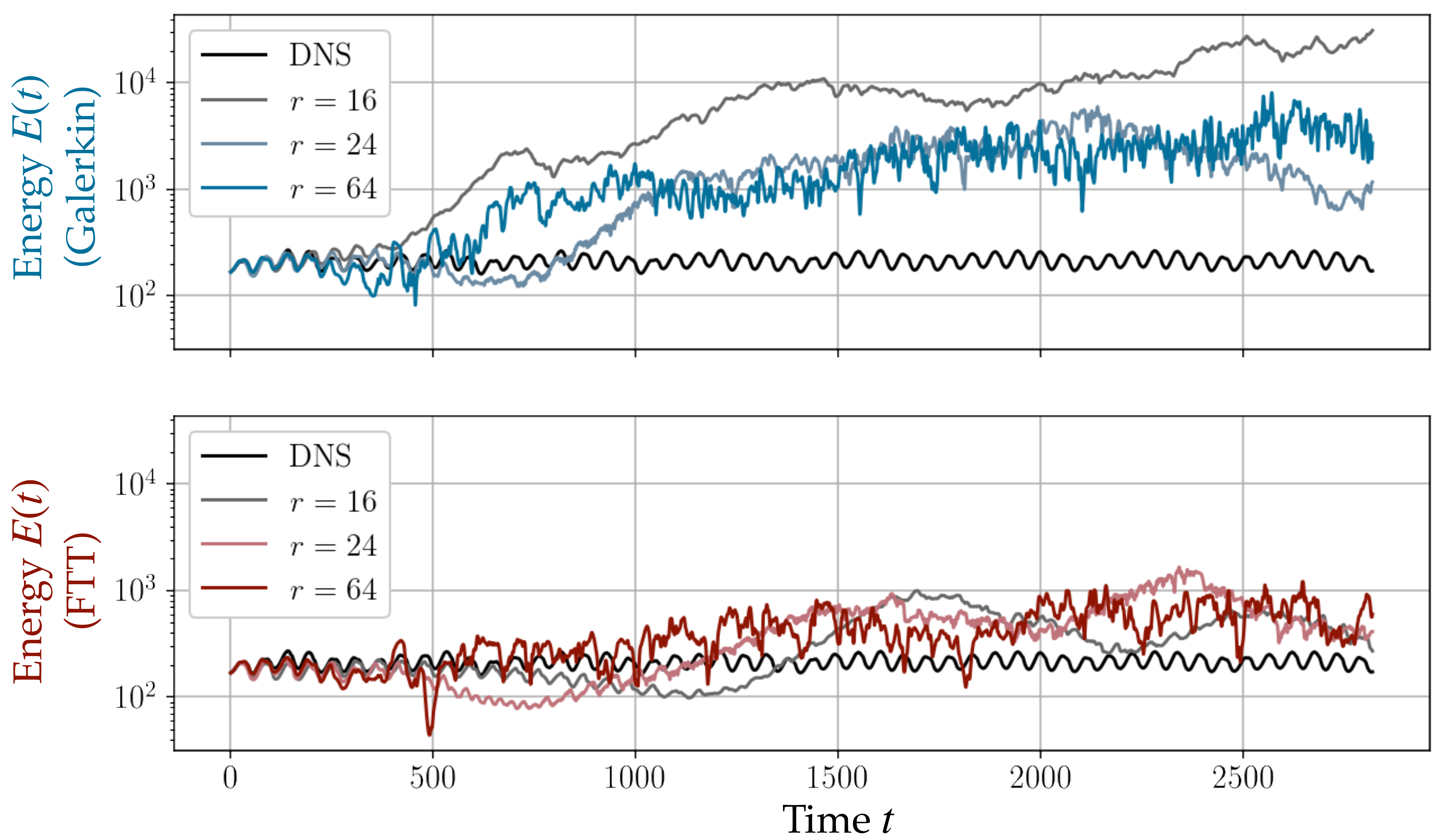}
\end{overpic}
\caption{
{\small
\textbf{Closure model based on the finite-time thermodynamics framework.}
The plots show the fluctuating kinetic energy in the first 16 modes of Galerkin (top) and FTT closure models (bottom) for the mixing layer on the longer domain.
Although the FTT closure model does stabilize the energy at a much lower level than the unmodified Galerkin system, it does not reproduce the natural dynamics of the flow over long time horizons.
} }
\label{fig: ftt -- results}
\end{figure}

The scalar eddy viscosity parameter $\nu_T$ is selected so that the mean value $\overline{K(t)}$ predicted by the model~\eqref{eq: ftt -- closed-galerkin} is consistent with the computed value of $K_\infty$ by the least-squares solution of the energy balance equation
\begin{equation}
    0 = \E \left\{ \sum_{i=1}^r  x_i f_i(\bm x) \right\} + \nu_T \E \left\{ \sum_{i=1}^r  L_{ii}^V x_i^2 \sqrt{\frac{K(t)}{K_\infty}} \right\}.
\end{equation}
With this approach the solution of the closure model is guaranteed to be bounded, with analytic estimates of the maximum kinetic energy growth~\citep{Cordier2013}.

Figure~\ref{fig: ftt -- results} compares the FTT-based nonlinear eddy viscosity model to the full POD-Galerkin system modeling the mixing layer on the longest domain.
While the FTT model is significantly more stable than the truncated Galerkin system, it neither matches the frequency content of the DNS over long time horizons nor remains phase-coherent as for the MMR model (Figure~\ref{fig: results -- mix-phase}).
This is likely because the nonlinearity introduced by the FTT model~\eqref{eq: ftt -- closed-galerkin} models transfer of energy to unresolved scales, but does not do so in a way that leads to the synchronizing coupled-oscillator model in the same way as the MMR closure.
The truncation values $r$ shown here are typical of any values tested for $r < 64$ for both the truncated POD-Galerkin system and the FTT closure.